# Gas filaments of the cosmic web located around active galaxies in a proto-cluster

H. Umehata[1,2], M. Fumagalli[3,4,5], I. Smail[3], Y. Matsuda[6,7], A. M. Swinbank[3], S. Cantalupo[8], C. Sykes[3,4], R. J. Ivison[9,10], C. C. Steidel[11], A. E. Shapley[12], J. Vernet[9], T. Yamada[13], Y. Tamura[14], M. Kubo[6], K. Nakanishi[6,7], M. Kajisawa[15], B. Hatsukade[2], and K. Kohno[2].

[1]RIKEN Cluster for Pioneering Research, 2-1 Hirosawa, Wako-shi, Saitama 351-0198, Japan

[2]Institute of Astronomy, School of Science, The University of Tokyo, 2-21-1 Osawa, Mitaka, Tokyo 181-0015, Japan

[3]Centre for Extragalactic Astronomy, Department of Physics, Durham University, South Road, Durham, DH1 3LE, UK

[4]Institute for Computational Cosmology, Durham University, South Road, Durham, DH1 3LE, UK

[5]Dipartimento di Fisica G. Occhialini, Università degli Studi di Milano Bicocca, Piazza della Scienza 3, 20126 Milano, Italy

[6]National Astronomical Observatory of Japan, 2-21-1 Osawa, Mitaka, Tokyo 181-8588, Japan

[7]Department of Astronomy, School of Science, SOKENDAI (The Graduate University for Advanced Studies), Osawa, Mitaka, Tokyo 181-8588, Japan

[8]Department of Physics, ETH Zurich, Wolfgang-Pauli-Strasse 27, 8093, Zurich, Switzerland

[9]European Southern Observatory, Karl-Schwarzschild-Str. 2, D-85748 Garching, Germany

[10]Institute for Astronomy, University of Edinburgh, Royal Observatory, Blackford Hill, Edinburgh EH9 3HJ, UK

[11]Cahill Center for Astronomy and Astrophysics, California Institute of Technology, MS 249-17, Pasadena, CA 91105, USA

[12]Department of Physics and Astronomy, University of California, Los Angeles, 430 Portola Plaza, Los Angeles, CA 90095, USA

[13]Institute of Space an Aeronautical Science, Japanese Aerospace Exploration Agency, 3-1-1, Yoshinodai, Chuo-ku, Sagamihara, Kanagawa, 252-5210, Japan

[14]Division of Particle and Astrophysical Science, Graduate School of Science, Nagoya University, Nagoya 464-8602, Japan

[15]Research Center for Space and Cosmic Evolution, Ehime University, Bunkyo-cho 2-5, Matsuyama 790-8577, Japan

*Correspondence to: hideki.umehata@riken.jp

## Abstract

Cosmological simulations predict the Universe contains a network of intergalactic gas filaments, within which galaxies form and evolve. However, the faintness of any emission from these filaments has limited tests of this prediction. We report the detection of rest-frame ultraviolet Lyman-α radiation from multiple filaments extending more than one megaparsec between galaxies within the SSA 22 proto-cluster at a redshift of 3.1. Intense star formation and supermassive black-hole activity is occurring within the galaxies embedded in these structures, which are the likely sources of the elevated ionizing radiation powering the observed Lyman-α emission. Our observations map the gas in filamentary structures of the type thought to fuel the growth of galaxies and black holes in massive proto-clusters.

Cosmological simulations of structure formation predict that the majority of gas in the intergalactic medium (IGM) is distributed in a cosmic web of sheets and filaments, as a consequence of gravitational collapse (*1*). The intersections of these structures become the locations at which galaxies and their supermassive black holes (SMBHs) form and evolve (*2*). Streams of cool gas falling along IGM filaments, driven by gravity, are predicted to provide most of the gas required for the growth of galaxies and SMBHs (*3,4*). Direct detection of the cosmic web in the early Universe would allow tests of these predictions, both for the large-scale structure, and the formation and evolution of galaxies.

Galaxy formation models predict that at a redshift ($z$) of about 3, >60% of all gas in the Universe resides in filaments (*5*), However, their low density makes them difficult to observe in emission. Absorption spectroscopy using background sources, such as quasars, has been the primary method used to trace neutral hydrogen (H I) in the IGM (*6,7*), which has provided insights into the nature of the cosmic web (*6*). Nevertheless, it has not been possible to obtain a detailed picture of these filaments, as information is limited to one dimension along the line of sight to the background source. The low sky density of sufficiently-bright background sources prevents study of the cosmic web on scales finer than a few megaparsecs (Mpc) (*7*).

Imaging the cosmic web in emission would provide two dimensional information. Filaments are predicted to emit the hydrogen Lyman-α (Ly α) line via fluorescence induced by the ultraviolet background (UVB) radiation (*8*). The intrinsically low intensity of the UVB means the expected surface brightness of a filament is $\sim 2.5 \times 10^{-20}$ erg $s^{-1}$ $cm^{-2}$ $arcsec^{-2}$ at $z\sim3$ (*9*), so direct detection of UVB-induced fluorescent emission from IGM filaments has remained elusive (*10*). To circumvent this limitation, we examine regions where local ionizing sources, such as star-forming galaxies and/or active galactic nuclei (AGNs), boost the local radiation field and hence elevate the Ly α emission to detectable levels (*11*). Extended (up to hundreds of kiloparsecs) Ly α nebulae have been observed around quasars, with morphologies and kinematics suggestive of cosmic web filaments connecting to the quasar host galaxies (*12-15*). Similarly, by using Ly α emitting galaxies (LAEs) or extended emission arising from the circumgalactic medium (CGM) as tracers, statistical evidence for filaments has been reported (*16-18*). These studies do not directly connect the cosmic web to the population of galaxies and SMBHs on cosmological scales.

We searched for extended filamentary structures using the Multi Unit Spectroscopic Explorer (MUSE) on the European Southern Observatory's Very Large Telescope (VLT). Our observations targeted the galaxy proto-cluster SSA 22 at $z$=3.09 (*19*), which was already known to host a three-dimensional filamentary structure as traced by LAEs on a scale of 30 comoving Mpc (note that comoving distance remains constant with epoch if the two objects are moving with the Hubble flow; *20*). At the intersection of this large-scale structure lies the proto-cluster core, where several intensely star-forming galaxies are known to lie within a $2' \times 3'$ region around the core, which was previously mapped at 1.1 mm with the Atacama Large Millimeter/submillimeter Array (ALMA) (the ALMA Deep Field in SSA 22 (ADF 22)) (*21*). To trace extended Ly α emission in this region, we mapped the ADF 22 field with a six-pointing MUSE mosaic covering $116'' \times 169''$, equivalent to $0.9 \times 1.3$ physical Mpc at $z$=3.09 (Fig. 1, *22*).

We searched the MUSE data cube for extended Ly α emission in conjunction with a narrow-band image covering the expected wavelength of redshifted Ly α emission, taken with Suprime-Cam on the Subaru telescope (*22*). We identified extended Ly α emission with surface brightness $\Sigma_{\mathrm{Ly}\,\alpha}$>~ $3 \times 10^{-19}$ erg $s^{-1}$ $cm^{-2}$ $arcsec^{-2}$ across the observed field, visible in the optimally-extracted Ly α map (Fig. 2, *22*). This map shows bright areas associated with the CGM of galaxies, along with several patches of emission at low surface brightness that connect to, but are not immediately associated with, individual galaxies in this region. Most of this low surface brightness Ly α emission forms two main filaments, running in a north-south direction, each with a total extent of > 1 physical megaparsec in projection. The scale of this emission far exceeds the expected size of the dark matter halo of even the most massive individual galaxies at this epoch (the halo radius is ~100 kiloparsecs, for a $10^{12.5}$ $M_\odot$ halo at $z$~3, where $M_\odot$ is the mass of the Sun), so the Ly α signal likely traces a structure connecting several galaxies. This network of filaments likely extends beyond the region we mapped, because the Ly α emission is detected up to the edge of the MUSE field of view. As shown in Fig. 3A and 3B, the majority of the Ly α emission is detected over a line-of-sight velocity range between ~ $-500$ and ~ $+1000$ km $s^{-1}$ relative to $z$=3.09. This velocity range reflects not only the 3D distribution of matter on large scales, but also the gas kinematics within the proto-cluster core which, coupled with radiative transfer effects, can produce velocity gradients of several hundreds of kilometers per second.

Observations of the SSA 22 proto-cluster have detected 35 Ly α blobs (LABs), defined as extended Ly α nebulae with sizes between several tens and several hundreds of kiloparsecs (*23-25*). Two of them lie within our field of view, each with sizes of ~ 40 kiloparsecs when measured at a Ly α surface brightness threshold of $\Sigma_{\mathrm{Ly}\,\alpha}$=

$2.2 \times 10^{-18}$ erg $s^{-1}$ $cm^{-2}$ $arcsec^{-2}$ (*24*). Fig. 2 shows these two LABs are parts of a larger network of megaparsec-scale filaments. Embedded in these filaments are also other patches of enhanced Ly α emission, some of which are associated with galaxies. We interpret the previously known LABs as bright knots within a wider network of gas filaments, and surmise that the fainter and more extended Ly α-emitting gas in these filaments has previously eluded detection because of its low surface brightness (*18, 26*).

To explore the link between these filaments and the associated galaxy population, we have measured redshifts of galaxies in this field using a multi-wavelength spectroscopic data set. The deep 1.1 mm ADF 22 map enables us to identify submillimeter galaxies (SMGs), which are massive, intense starburst galaxies with large amounts of gas and dust in their interstellar mediums. X-ray luminous AGNs, which host growing SMBHs, were identified from observations using the *Chandra* space telescope (*22, 27*). Spectroscopic redshifts for these populations were determined using a combination of emission lines (CO $J$=3→2, H β, and [O III] 4959 & 5008 Å lines) from ALMA data and observations with the near-infrared multi-object spectrograph, Multi-Object Spectrometer for Infra-Red Exploration on the Keck I telescope (*22*). We confirm that 16 SMGs and 8 X-ray luminous AGNs are proto-cluster members, with redshifts $3.085 \le z \le 3.098$ (Table S2). All of the SMGs and X-ray luminous AGNs are distributed within the same structure (see Figs. 2 & 3), closely tracking the Ly α filaments both spatially (in projection) and in velocity (Fig. S10). A similar pattern is also evident for normal star-forming galaxies, and LAEs (Fig. S7). We interpret this close alignment as evidence that the Ly α filaments are directly linked to the population of active galaxies and SMBHs. Gas filaments are thought to supply (under the effect of gravity) the fuel for active SMGs and X-ray bright AGNs.

The filaments have Ly α brightness above the level expected from fluorescent emission induced by the UVB (*8, 11*). Radiative transfer calculations predict a maximum surface brightness from optically-thick gas of ~$2.5 \times 10^{-20}$ erg $s^{-1}$ $cm^{-2}$ $arcsec^{-2}$ assuming a $z$~3 UVB (*9*). Our observations contain emission at levels of $\Sigma_{\mathrm{Ly}\,\alpha} \ge 3 \times 10^{-19}$ erg $s^{-1}$ $cm^{-2}$ $arcsec^{-2}$, which in the optically-thick limit requires an intensity of the ionizing radiation field that is > 12 times brighter than predicted by UVB models (*22*). This corresponds to an ionizing photon flux $\phi > 4 \times 10^{6}$ $cm^{-2}$ $s^{-1}$. If we assume the gas is fully ionized, the observed surface brightness would imply even higher photon fluxes, and densities of $\sim 6 \times 10^{-3}$ $cm^{-3}$ (*12*) for gas at $T$ ~ $10^4$ K, where $T$ is temperature, and a typical filament width of 100 kpc (Fig. 2). Fluctuations in the observed surface brightness suggest a variable density across the structure, as commonly found for bright Ly α nebulae (*12*). Fig. 2 also shows regions at much higher surface brightness, particularly overlapping with galaxies (see also Fig. S7). The fainter emission regions display lower velocity dispersions (the full width at half maximum (FWHM)~150 km $s^{-1}$) than the brighter knots (FWHM~730 km $s^{-1}$), where the surface brightness of the latter is comparable to the typical brightness of the LABs (Fig. S4 and Fig. S9). This higher surface brightness may indicate the presence of localized sources of ionization, such as star formation or AGN activity, as commonly seen in LABs (*28*).

Given the large number of active galaxies in this region, the elevated photon flux required to power the filaments could be provided by the galaxy population identified within our field. To test this hypothesis, we evaluate number of ionizing photons provided by X-ray selected AGNs and SMGs (*22*). Under the simple assumption that the ionizing sources typically lie 250 kpc from filaments, the required photon flux corresponds to a photon number emission rate $Q_{\rm ion}$ ~$10^{55}$ $s^{-1}$ as a whole. The eight X-ray AGNs in the structure have $L$x~$10^{44}$ erg $s^{-1}$, corresponding to a total rate of $Q_{\rm ion}$ ~$10^{57}$ $s^{-1}$, while the 16 SMGs which are proto-cluster member are forming stars at a rate of 160-1700 $M_{\odot}$yr$^{-1}$ and hence produce a total of $Q_{\rm ion}$ ~$10^{57}$ $s^{-1}$. This is sufficient ionizing photon flux to power the filament emission, even under our assumption that only 1% of the photons escape their host galaxies. This simple estimate, although an approximation of the more complex radiative transfer in this region, supports our interpretation that the gas residing in these filamentary structures is ionized by the photons produced by star-forming galaxies and AGNs in the massive proto-cluster core.

The volume densities of SMGs and X-ray AGNs in this field are about three orders magnitude higher than the volume average at this epoch (*21*). Such an overdensity of active populations is very rare (*29*), and there is little observational evidence regarding how this intense activity is fueled and sustained. Cosmological simulations suggest that rapid infall of gas from the cosmic web in proto-clusters may lead to the formation of SMGs (*30*). While gas inflows are not directly observable in our data, the location of SMGs and AGNs within the filaments supports the idea that large reservoirs of gas are funneled towards forming galaxies under the effect of gravity, triggering and sustaining their star formation and driving the growth and activity of their central SMBHs. Assuming a typical density of $6 \times 10^{-3}$ cm$^{-3}$ for filaments with (projected) thicknesses of ~100 kiloparsecs, the region imaged by our observations contains ~ $10^{12}$ $M_{\odot}$ of gas (depending on the filling factor of the gas), which is potentially available to accrete onto galaxies in this region and so fuel their continuing star formation (*22*).

Our observations have uncovered a large-scale filamentary structure in emission in the core of the SSA 22 proto-cluster. Evidence of similar structures in other proto-clusters from imaging observations (*15*, *25*) suggests that this may be a general feature of proto-clusters in the early Universe. The network of filaments in SSA 22 is found to connect individual galaxies across a large volume, allowing it to power star formation and black hole growth in active galaxy populations at $z$~3.

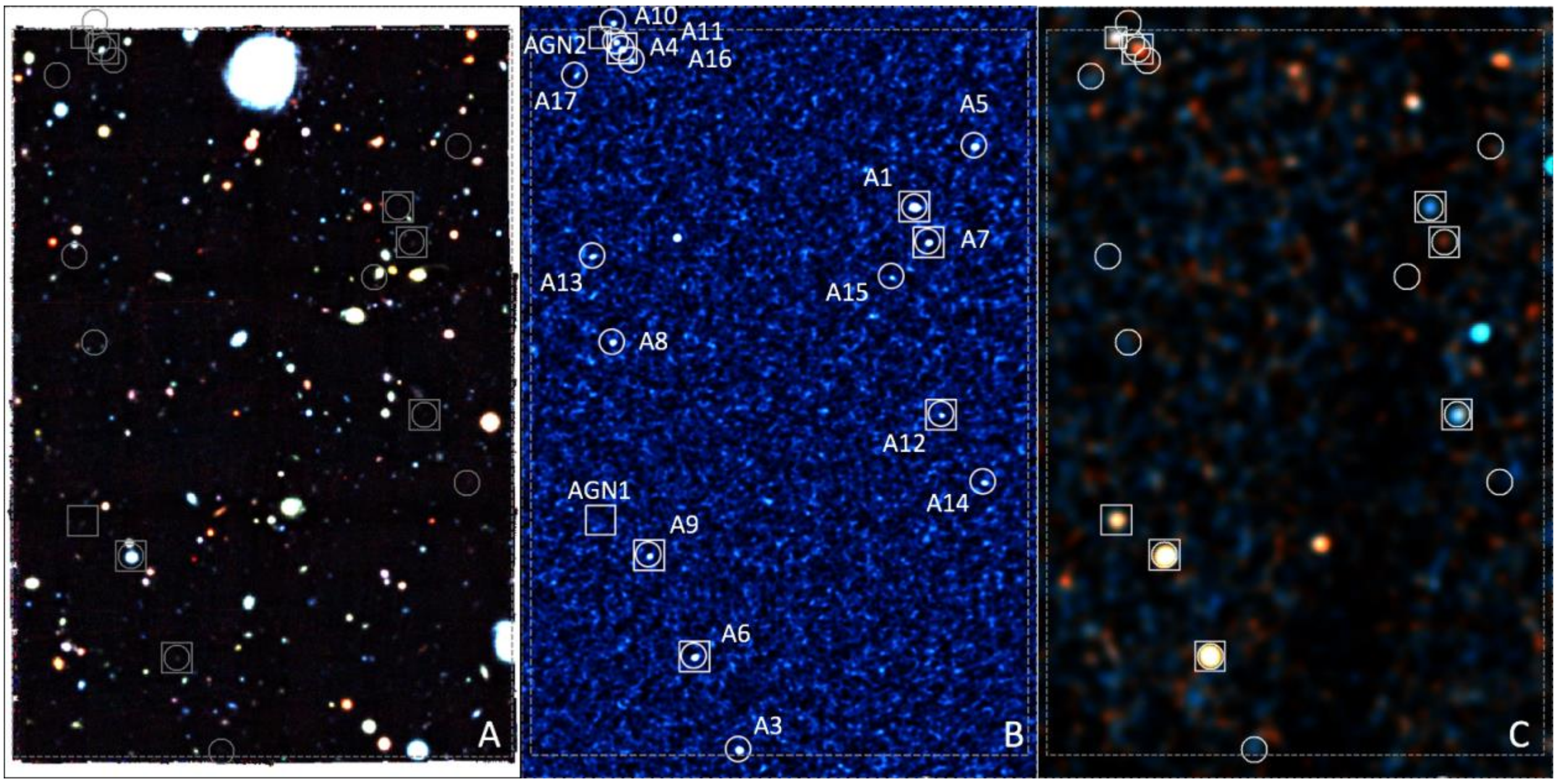

**Fig. 1. Multi-wavelength images of ADF 22, illustrating the overdensity of galaxies and AGNs in a narrow redshift range at *z*=3.09**. Each panel is centered at (α, δ) = (22h17m34.0s, +00d17m00s) (where α is right ascension and δ is declination), and $2' \times 3'$ in size, with the inner dashed area showing the MUSE coverage, $116'' \times 169''$ (0.9 × 1.3 Mpc at *z*~3.1). North is up and east is left. **(A)** A pseudo-color map created from the MUSE cube (synthetized *V*-, *R*-, and *i'*- bands are used for the blue, green, and red channels). **(B)** The 1.1-mm ALMA continuum map of ADF 22 (*22*). Identified sources at *z*=3.09 are marked with white circles (SMGs) and squares (AGNs); positions and redshifts are listed in Table S2. **(C)** A pseudo-color map of the *Chandra* X-ray data. 2-8 keV (hard band), 0.5-8 keV (full band) and 0.5-2 keV (soft band) are utilized for the blue, green, and red channels.

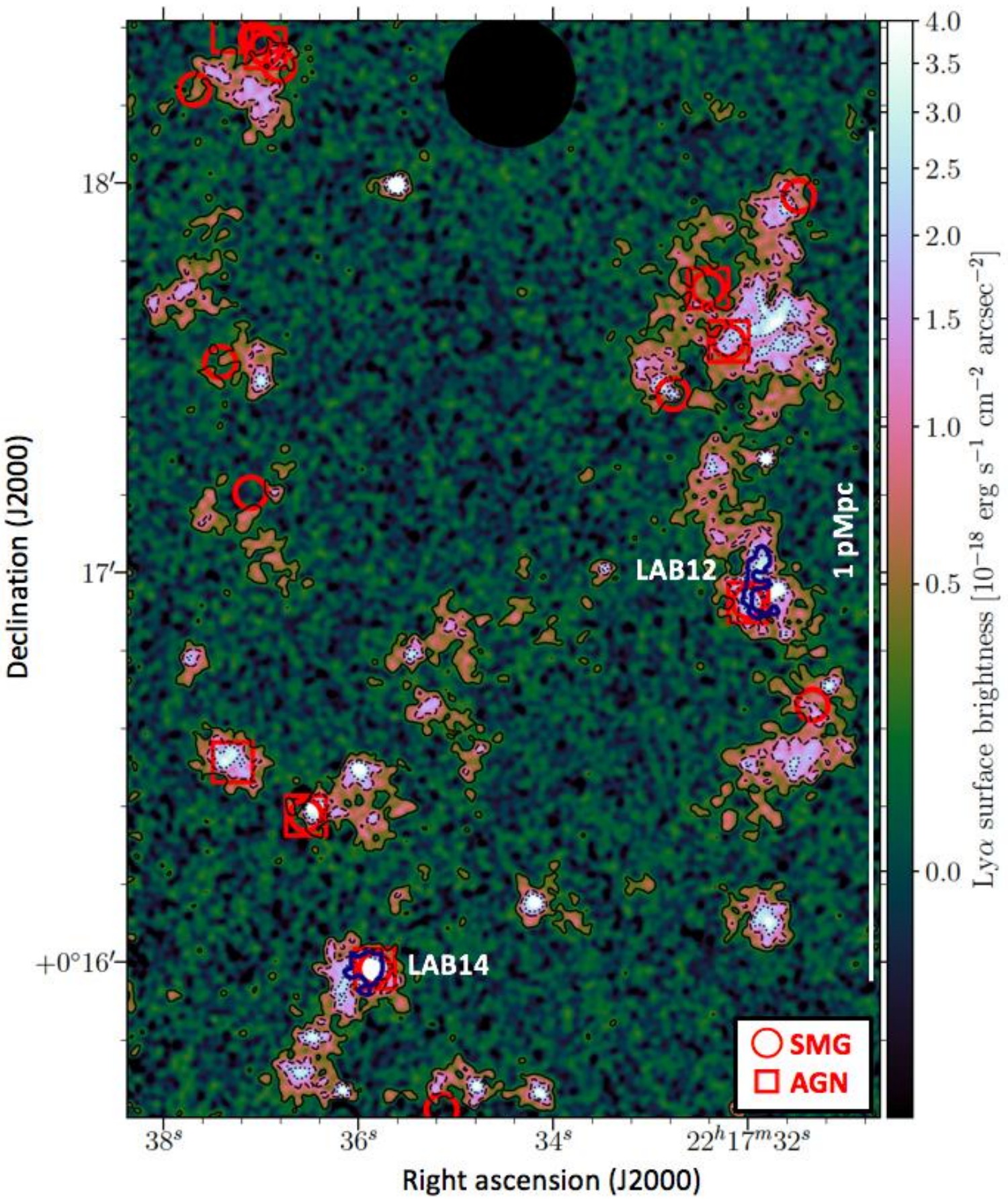

**Fig. 2. Ly α emission map optimally extracted from the MUSE observations, covering the same field as Fig. 1.** Ly α emissions largely compose two groups of filamentary structures for more than 1 physical Mpc. One high surface brightness filament is visible running North-South on the west side of the field, while a fainter (and hence more fragmented) structure runs North-South up the east side of the field. Contours with solid, dashed, and dotted lines show Ly α surface brightness levels of $\Sigma_{Ly\,\alpha}$ = 0.3, 1.0, and $2.0 \times 10^{-18}$ erg $s^{-1}$ $cm^{-2}$ $arcsec^{-2}$, respectively (these correspond to $2\sigma$, $7\sigma$, and $13\sigma$ above the representative noise level). Navy contours indicate the extent of the two LABs in this field (*23*). Positions of SMGs and X-ray luminous AGNs at *z*=3.09 are also shown. The large, filled black circle shows data removed around a foreground a low-redshift galaxy.

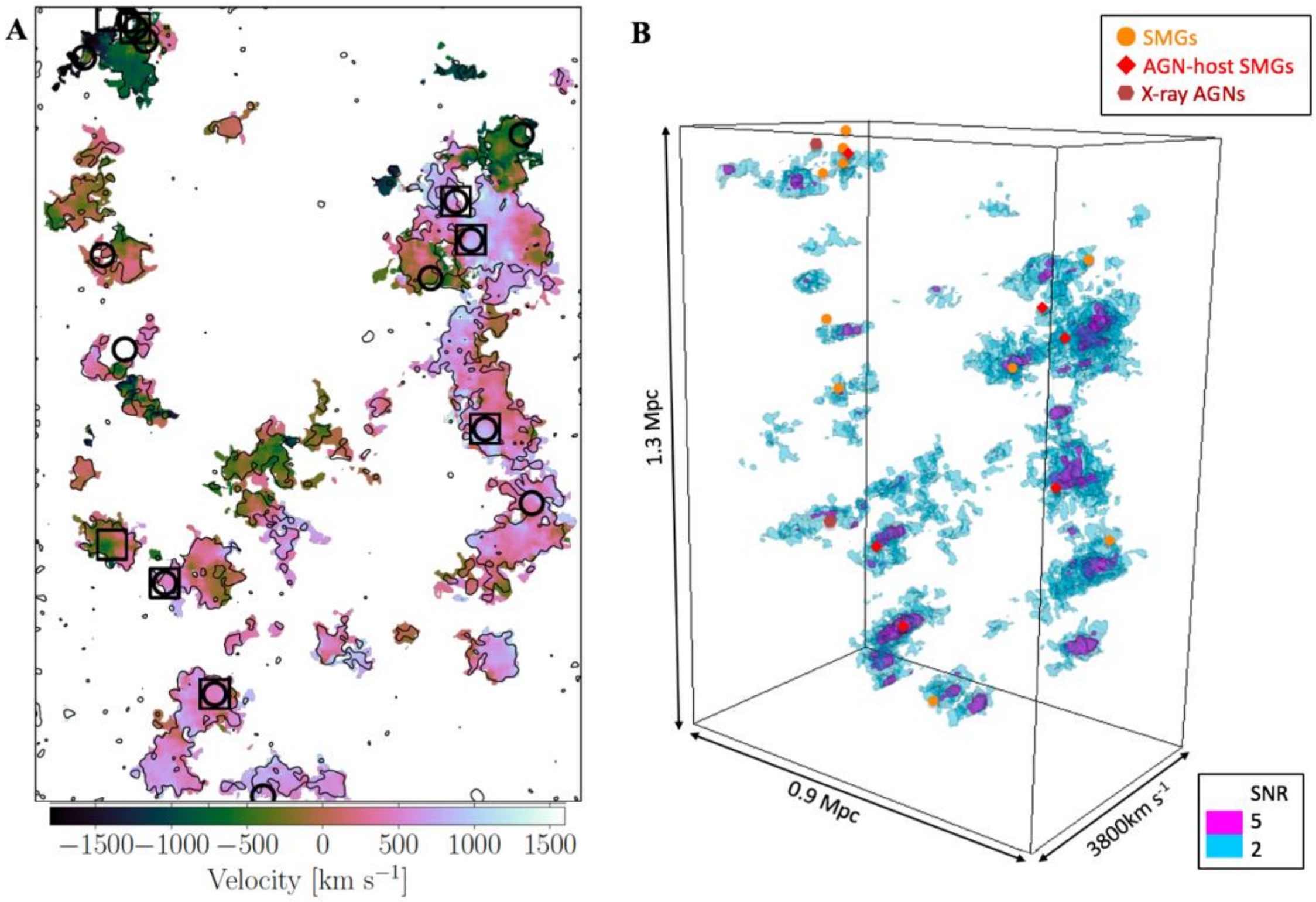

**Fig. 3. The three-dimensional pictures of Ly α filaments. (A)** Velocity map of the Ly α emission, obtained from its flux-weighted-centroid in the MUSE data . Image scale and plotting symbols are the same as Figure 2. Coherent velocity trends can be seen along the filament structures. **(B)** The three-dimensional distribution of Ly α filaments, shown with blue (SNR>2, where SNR means signal-to-noise ratio) and magenta (SNR>5) voxels. The locations of SMGs (without detectable X-ray AGNs, orange circles), AGN-hosting SMGs (red diamonds), and X-ray bright AGNs without ALMA 1mm detections (brown hexagons) are also displayed. The Ly α filaments and SMGs/AGNs are co-located on megaparsec scales.

**Acknowledgments:**
We thank the reviewers for their constructive comments, which was very helpful to improve this paper. Based on observations collected at the European Organisation for Astronomical Research in the Southern Hemisphere. We thank all ESO staff who supported us in observation preparation and execution.. We thank JAO and EA-ARC staffs for the preparation, observation, and initial data reduction. ALMA is a partnership of ESO (representing its member states), NSF (USA) and NINS (Japan), together with NRC (Canada) and NSC and ASIAA (Taiwan) and KASI (Republic of Korea), in cooperation with the Republic of Chile. The Joint ALMA Observatory is operated by ESO, AUI/NRAO and NAOJ. Some of the data were obtained at the W. M. Keck Observatory, which is operated as a scientific partnership among the California Institute of Technology, the University of California and the National Aeronautics and Space Administration. We thank Sherry Yeh for her assistance on the MOSFIRE observations. The observatory was made possible by the generous financial support of the W. M. Keck Foundation. The authors wish to recognize and acknowledge the very significant cultural role and reverence that the summit of Mauna Kea has always had within the indigenous Hawaiian community. We are most fortunate to have the opportunity to conduct observations from this mountain.

**Funding:**

H.U., Y.M., B.H., Y.T., and K.K. are supported by JSPS KAKENHI [grant number 17K14252, 25287043/17H04831/17KK0098, 19K03925, 17H06130, and 17H06130, respectively]. Y.T. acknowledges support from NAOJ ALMA Scientific Research Grant Numbers 2018-09B. M.F., I.R.S. and A.M.S. acknowledge support by the Science and Technology Facilities Council [grant number ST/P000541/1]. This project has received funding from the European Research Council (ERC) under the European Union's Horizon 2020 research and innovation programme (grant agreement No 757535). C.S. acknowledges an STFC studentship [grant number ST/R504725/1]. S.C. gratefully acknowledges support from Swiss National Science Foundation grant PP00P2_163824.

**Author contributions:**

All authors meet the journal’s authorship criteria. H.U. led the project and analyzed the data (MUSE, Subaru, ALMA, Keck). M.F., I.R.S., S.C. and A.M.S. worked on the MUSE data reduction. C.C.S. worked on the Keck data reduction. Y.M. worked on the Subaru data reduction. M.F. and C.S. performed radiative transfer calculations. R.J.I., A.E.S, J.V., T.Y., Y.T., M.K., K.N., M.K., B.H., and K.K. contributed to interpret the results. All authors reviewed, discussed, and commented on the results and on the manuscript.

**Data and materials availability:**

The MUSE data are available through the ESO archive (http://archive.eso.org/cms.html) under programs 099.A-0638 and 0101.A-0679, the ALMA data through the archive (https://almascience.nrao.edu/ alma-data/ archive) under ADS/JAO.ALMA programs #2013.1.00162.S, #2015.1.00212.S and #2016.1.00543.S, and MOSFIRE observations through the Keck Observatory Archive (https://www2.keck.hawaii.edu/ koa/ public/ koa.php) under programs 2017B_S412.

**Supplementary Materials:**

Data, Methods, and Supplementary Text S1-S11

Figures S1-S10

Tables S1-S2

References *(31-76)*

Data S1-S2

# Supplementary Materials for

## Gas filaments of the cosmic web located around active galaxies in a proto-cluster

H. Umehata*, M. Fumagalli, I. Smail, Y. Matsuda, A. M. Swinbank, S. Cantalupo, C. Sykes, R. J. Ivison,

C. C. Steidel, A. E. Shapley, J. Vernet, T. Yamada, Y. Tamura, M. Kubo, K. Nakanishi, M. Kajisawa,

B. Hatsukade, and K. Kohno.

*Correspondence to: hideki.umehata@riken.jp

**This PDF file includes:**

Data, Methods, and Supplementary Text S1 to S11

Figs. S1 to S10

Tables S1 to S2

# S1 Cosmology

Throughout this paper, we use a Λ Cold Dark Matter cosmology with Hubble constant $H_0$=70 km $s^{-1}$ $Mpc^{-1}$, matter density parameter $\Omega_m$=0.3 and dark energy density parameter $\Omega_\Lambda$=0.7. This produces a physical scale of 7.6 kpc $arcsec^{-1}$ at $z$=3.09.

# S2 Target Information

Our target field is a sub-region of the $z$=3.1 SSA 22 proto-cluster core. A filamentary structure spanning 50 comoving Mpc has been found using LAEs as a tracer in both two-dimensions (Fig. S1, (*31*)) and three-dimensions (*20*). Toward the proto-cluster core, which is interpreted as an intersection of this three-dimensional filamentary structure, an ALMA 1.1 mm map has been obtained to identify dust-obscured star-formation activity and named as ADF 22 (*21, 32, 33*). The approximate $2' \times 3'$ field centered at (α, δ) = (22h17m34.0s, +00d17m00s) has a typical r.m.s. (root mean square) level of 60 $\mu$Jy $beam^{-1}$ with $0''.7$ angular resolution (*32*). In total, 18 SMGs, which we denote ADF 22. A1 to ADF 22. A18, with flux density $S_{1.1mm}$ = 0.5-5.6 mJy, are found in the field (Fig. 1).

The ADF 22 field is fully covered by a deep *Chandra* X-ray survey (Fig. 1C) (*27,34*). In total, 19 X-ray sources are identified in the $2' \times 3'$ region (*34*). Among them, eight X-ray sources have been identified as X-ray luminous AGNs at $z$~3 (*21, 27, 28, 33, 35*). These X-ray sources are classified as an AGN on the basis of X-ray luminosity ($L_X$>$3 \times 10^{42}$ erg $s^{-1}$; e.g., *36*). For the remaining 11 X-ray sources, 9 have been found to be at lower redshift (e.g., *34*, *37*). Two sources have no counterpart at any other wavelengths except for X-ray, and hence there is no constraint on their redshift or physical properties.

# S3 Suprime-Cam Data Reduction and Analysis

ADF 22 has been observed with a narrow-band image taken with the Subaru/Suprime-Cam (*38*) using the NB497 filter to trace Ly α emission at z~3.09. The NB497 filter has a central wavelength of 4977 Å and a full width at half maximum of 77 Å, covering Ly α in the range $z$=$3.06 - 3.13$ (*31*). The narrow-band image was used to map the distribution of LAEs and LABs in this field (*23, 31*). As described in the main text, two LABs, LAB 12 and LAB 14, are found in the region of ADF 22 (*24*) (one LAB candidate, LAB 36, has also been reported (*35*)).

We used the narrow-band image to search for extended Ly α structures in ADF 22. To identify faint extended emission, we produced a smoothed narrow-band image with a $4'' \times 4''$ (or 20×20 pixels, given the Suprime-Cam pixel scale of 0.2 arcsec) median filter, masking bright continuum sources. A noise level of $3.4 \times 10^{-19}$ ergs $s^{-1}$ $cm^{-2}$ $arcsec^{-2}$ ($1\sigma$) was obtained, as measured within 10000 randomly distributed apertures with a diameter of $4''$. The Fig. S2A shows the resulting image. We extracted extended Ly α structures which have areas of more than 48 $\mathrm{arcsec}^2$ (2800 $\mathrm{kpc}^2$ or 1200 connecting pixels) above $\Sigma_{\mathrm{Ly}\alpha} = 0.3 \times 10^{-18}$ ergs $\mathrm{s}^{-1}$ $\mathrm{cm}^{-2}$ $\mathrm{arcsec}^{-2}$ using SEXTRACTOR (v2.5.0; *39*) as in previous work (*24*). As shown in Fig. S2A, this analysis suggests the presence of Ly α filaments that extend to large scales, and which have not been identified in previous studies (*24*). The extent of these structure reaches approximately one megaparsec (the map shows an area of $0.9 \times 1.3$ physical Mpc at $z$=3.09). To confirm this with independent observations, we followed-up this emission using MUSE, which also enables us to determine the velocity structure of this emission.

# S4 MUSE Observations, Data Reduction, and Analysis

We used the MUSE instrument at the UT4 VLT (*40*) to observe a sub-region of ADF 22. Considering the field of view of MUSE in the Wide Field Mode (WFM, $1' \times 1'$), we acquired 6 adjacent fields with a $2 \times 3$ pattern. Each pointing overlaps with the adjacent pointing by $5''$ in Right ascension and Declination to obtain contiguous coverage. After mosaicking, this configuration yielded a total field of view of $116'' \times 169''$ centered at (α, δ) = (22h17m34.5s, +00d17m00.5s) with position angle 0 degree. To obtain a full mosaic with uniform sensitivity, we adopted a total on-source exposure time of 4.17 hours for each individual tile, each divided into ten 1500 sec individual exposures. Each exposure was rotated by 90 degrees with respect to the previous one, and slightly shifted to reduce systematic noise. To maximize the sensitivity at wavelengths shorter than 5000 Å (which contains our target emission line), we used the extended mode without adaptive optics. Our observations were initiated in 2017. Out of the 30 hours including overheads awarded (programme number 099.A-0638, PI. H. Umehata), 8/30 hours (27%) of observations were completed between June and September 2017. The remaining 22/30 hours (73%) of observations were completed between June and September 2018 (programme 101.A-0679, PI. H. Umehata). The weather conditions were generally classified as clear sky, but with some non-photometric conditions. The seeing was typically 1 arcsec, with a range between 0.6 arcsec to 1.7 arcsec, according to the weather data recorded with the observing blocks. We adopted the standard calibration for both programmes.

We used the standard ESO MUSE pipeline (v2.4.1) for the basic data reduction (*41*). This includes bias subtraction, flat-fielding, twilight and illumination correction, and wavelength calibration for each of the individual exposures. A spectrophotometric standard star observed during the same night with the same configuration was utilized for the flux calibration of each exposure. Sky-subtraction, necessary to detect faint and extended emission, was not carried out with the ESO pipeline due to known residuals arising when fitting sky lines. Instead, we used the CUBEXTRACTOR package (v.1.8) (*14,42*). Following the basic reduction with the ESO pipeline as described above, this package performs an additional correction to homogenize the illumination across the field and as a function of wavelength, and then performs sky subtraction with better accuracy than the MUSE pipeline. This post-processing is necessary to extract low surface brightness emission from MUSE data cubes (e.g. (*14*)).

To construct the final mosaic, one of the exposures was selected as a reference frame relative to which additional overlapping exposures were aligned. The final astrometric solution of the mosaic was derived from a Subaru/Suprime-Cam *R*-band image (*31*). A single calibrated science data cube was produced, together with a variance cube that reflects the propagation of uncertainties associated with the MUSE pipeline and the combination of different exposures. However, due to correlated noise at pixel level, this variance cube underestimates the true uncertainties, and we therefore rescale the variance computed by the pipeline to match the empirical variance derived in the final cube as a function of wavelength. For this, we first compute the standard deviation of the distribution of fluxes normalized by the pipeline uncertainty in each layer. Given that the pipeline uncertainty is underestimated, these standard deviations are greater than unity, and have a value equal to the correction needed to match the uncertainty to the data r. m. s. We therefore fitted a smooth spline function to these values, and used it to correct the pipeline standard deviation. A pseudo-color image from the final MUSE cube is shown in Fig. 1A, using a set of reconstructed broad band images (5000-6000 Å, 6000-7000 Å, and 7500-8500 Å as a proxy of *V*-, *R*-, and *i*'-band, respectively).

Before searching for low surface brightness emission in the cube, we subtracted the image of a quasar at $z$=3.09 (SSA 22 a D13) (*19*) by fitting its point spread function (PSF) using the CUBEXTRACTOR/CUBEPSFSUB algorithm (another source, the quasar ADF 22 A9, has been also detected at 1.1 mm and hence it is one of our SMGs in the proto-cluster (*21*)). During this process, we empirically obtained a PSF model from the quasar light profile on pseudo-NB images as a function of wavelength. MUSE PSFs are stable during long exposures (e.g., *43*). We selected a subset of the cube spanning 150 spectral pixels, or 187.5 Å, within which we mask the wavelength range where the redshifted Ly α emission line was expected (4940.4 Å-5002.9 Å) (see (*44*) for more details). After PSF subtraction, we proceeded to subtract the continuum of the remaining sources using the CUBEBKGSUB procedure in CUBEXTRACTOR, which is based on a fast median-filtering approach. Again, we masked the possible range of wavelength where Ly α line emission from the sources at $z$~3.09 was expected.

Following this preparatory step, we searched for extended Ly α emission at $z$~3.09 in the MUSE cube in two ways. As the first approach, we created narrow-band images from the MUSE cube. This method has the advantage of enabling direct comparisons between the results from the MUSE and the Suprime-Cam observations. As in the case of the Suprime-Cam image, we applied a $4'' \times 4''$ (or $20 \times 20$ pixels) median filter, masking bright continuum sources identified on a white-band image from the MUSE cube to avoid contamination from residual continuum emission. The middle panel in Fig. S2 shows a 37.5 Å width narrow-band image. Emission with similar morphology is detected independently in both datasets. More quantitatively, we compared the MUSE narrow-band maps with the Suprime-Cam map in the right panel of Fig. S2. This panel shows the Ly α structure extracted from the Suprime-Cam image where Ly α emission is also detected by at least one of the MUSE narrow-band maps above the same threshold of $\Sigma_{\mathrm{Ly}\,\alpha} = 0.3 \times 10^{-18}$ ergs $s^{-1}$ $cm^{-2}$ $arcsec^{-2}$. This threshold corresponds to $2\sigma$ noise level measured, as in the case of Suprime-Cam map, using 10,000 randomly-distributed apertures with a diameter of 4". The majority of the Ly α structure identified in the Suprime-Cam image is re-identified in the MUSE narrow-band maps. The simultaneous detection with two independent instruments provides us with confidence in the robustness of the detected extended Ly α emission on megaparsec-scale within this field.

Having confirmed that extended low surface brightness emission is present across the field, we proceed to extract a Ly α map with a second method, which refines the narrow-band extraction in several ways. First, we may miss emission with relatively narrow velocity width because the significance of such line emission would decrease in narrow-band images which cover a wide range of velocity, as is especially the case of the Suprime-Cam image. Second, some of the emission identified in projection with narrow-band maps may be composed of distinct velocity components overlapping along the line of sight. Third, the low spectral resolution of narrow-band maps is insufficient to trace velocity structure. To obviate to these limitations, we searched for Ly α emission in three dimensions by selecting volume pixels (voxels) with line emission belonging to coherent structures.

To this end, we used CUBEXTRACTOR to select groups of voxels with connected extended emission within the cube, as done in previous work *(14, 42)*. The extraction algorithm relies on three parameters: the width of a Gaussian kernel used to smooth the data, the surface brightness threshold needed to identify each voxel, and a minimum number of connected voxels that identify a physical structure. In principle, a large kernel is favored to trace extended emission, but to guard against the possibility that noise in MUSE has unidentified non-gaussian behaviour, we adopt a more conservative approach to ensure we do not misidentify positive peaks in the noise as real emission. Moreover, while a large number of connected voxels is optimal to trace extended and coherent structure, one may miss real emission in more compact knots. To strike a balance between these competing effects, we adopt the following method.

First, we focus on relatively high signal-to-noise and potentially more compact Ly α emission within the MUSE cube. We applied a compact Gaussian filter with 3 pixel radius ($0''.6$, or 4.6 kpc) in the spatial direction, masking continuum sources (to avoid contamination from low amplitude residuals that may artificially inflate the signal at low surface brightness levels) and any residual artefact such as cosmic rays. We did not apply any smoothing in the wavelength direction to maximize the spectral resolution. As for the remaining two parameters, we adopted the following two choices: (i) a minimum of 4500 voxels above a SNR> 2 and (ii) more than 1500 voxels above SNR > 2.5. A three-dimensional segmentation map was then created, which identifies the voxels associated with the extended Ly α emission. Using this segmentation map, an optimally-summed 2D narrow-band image was reconstructed by integrating the flux in the 3D mask along the line of sight at each spatial pixel (spaxel). In this context, ‘optimally summed’ refers to a line map that is constructed by integrating flux in wavelength only within pixels inside the 3D segmentation map to maximize the signal-to-noise ratio by excluding voxels where no signal is detected. With this technique, the number of voxels utilized to make the 2D image differs between each spatial pixel. The resultant image, which we call the "compact layer" is shown in Fig. S3A. While we considered the actual noise level for each voxel in the extraction process, we also derived an empirical r.m.s. level of the smoothed cube by fitting a one-sided Gaussian to the distribution of negative fluxes. The derived noise level for the narrow-band map was $0.25 \times 10^{-18}$ ergs $s^{-1}$ $cm^{-2}$ $arcsec^{-2}$ $(1\sigma)$, assuming Ly α emission detected over three adjacent spaxels. The Ly α map obtained recovers a large part of the Ly α filaments identified by the narrow-band method described above.

After the extraction of this layer, we searched for fainter and more extended signal by means of a larger smoothing kernel. However, to overcome possible issues related to the marginally non-Gaussian noise, we proceed more conservatively, flagging as detections only signals that are detected both by MUSE and in the Suprime-Cam narrow-band image (we note that we did not used the Suprime-Cam information for the above-mentioned "compact layer".). To increase sensitivity to diffuse and extended Ly α emission, we adopted a Gaussian filter with spatial radii of 5 pixels (1", or 7.6 kpc), and a minimum of 5000 voxels above a signal-to-noise ratio > 2. In this case, the typical narrow-band sensitivity was estimated to be $0.15 \times 10^{-18}$ ergs $s^{-1}$ $cm^{-2}$ $arcsec^{-2}$ $(1\sigma)$ with similar calculation described above (see Fig. S3B for the flux distributions of voxels). After this initial extraction process on the MUSE data, we flagged as detections only voxels which are encompassed by the detected Suprime-Cam structures (Fig. S2), a conservative choice we make when working in this low-surface brightness regime. The surface brightness limit is $0.3 \times 10^{-18}$ ergs $s^{-1}$ $cm^{-2}$ $arcsec^{-2}$ as we described above, and hence both detection limits are roughly equivalent. The final resultant narrow-band image, combined with the compact layer, is shown in Fig. 2 and Fig. S3B (we refer to it as "compact + extended"). We use this cube and images for further analysis and discussion. The "extended" component is distinct from emission of continuum sources, as we find patches without the presence of galaxies, with a UV star formation upper limit of 0.7 $M_{\odot}$ $yr^{-1}$ $arcsec^{-2}$ $(3\ \sigma)$. The detected extended emission is unlikely to be a blend of several LAEs, given that sources with a star-formation-rate (SFR) of ~0.1 $M_{\odot}$ $yr^{-1}$ and size of 2-3 kpc would be detected as a distinct knot of emission.

We measured the total extent, integrated flux, averaged surface brightness, and integrated luminosity of the extended Ly α emission using the reconstructed MUSE narrow-band maps (including the compact layer only map for comparison). For the two MUSE maps, we adopt the $2\sigma$ surface brightness limits ($0.3 \times 10^{-18}$ ergs $s^{-1}$ $cm^{-2}$ $arcsec^{-2}$), and measured properties for emissions above this threshold. The results are summarized in Table S1, together with the properties of two LABs located in the field *(23)*. We also performed Ly α line-stacking analysis to derive average line velocity profiles at different levels of surface brightness. Because Ly α is a resonant line which often shows an asymmetric profile, we adopt the flux-weighted mean as a central velocity for a given spatial pixel to evaluate the averaged line profile. For the Ly α filaments, we stacked the line emission shifting spectra relative to the flux-weighted mean, to remove differences in the line of sight velocity across the map. The stacked spectra were derived for three ranges of projected surface brightness: $(0.3-1.0)\times 10^{-18}$ ergs $s^{-1}$ $cm^{-2}$ $arcsec^{-2}$, $(1.0-2.0)\times 10^{-18}$ ergs $s^{-1}$ $cm^{-2}$ $arcsec^{-2}$, and $(2.0-6.0)\times 10^{-18}$ ergs $s^{-1}$ $cm^{-2}$ $arcsec^{-2}$. The results are shown in Fig. S4. In the case of the faintest emission, fitting with a single Gaussian shows that FWHM of the composite line is $144\pm 4$ km $s^{-1}$. This indicates that the assumption we adopted in evaluating the sensitivity limit on the reconstructed narrow-band map using three adjacent spaxels (~225 km $s^{-1}$) is valid. The stacked spectra show a clear line profile, supporting an astrophysical origin for the extracted low-surface brightness signal.

# S5 ALMA Band3 Data and Analysis

Dust extinction makes it challenging to detect rest-frame UV emission lines from SMGs, but they often have abundant molecular gas reservoirs so exhibit bright molecular gas line emission (*45*). We can therefore exploit molecular line searches to measure redshifts of the SMGs detected within the footprint of our MUSE observations. We observed the majority of the region of ADF 22 using ALMA band 3 as part of two programmes (IDs. 2015.1.00212.S, 2016.1.00543.S; PI. H. Umehata) to detect redshifted CO ($J$=3→2) ($\nu_{rest}$=345.796 GHz) line at $z$~3.09. The observations were performed in two periods, from July 21st to August 19th, 2016, and from May 7th to July 4th, 2016 with the same observational set-up. We used the frequency-division mode (FDM) correlator in which four spectral windows have a bandwidth of 1.875 GHz and a spectral resolution of 977 kHz. The four spectral windows had central frequencies of 85.04 GHz, 86.87 GHz, 97.03 GHz, and 98.87 GHz, to observe the CO line at z=2.47-2.60 and 2.94-3.11. A 13-point mosaic centered at (α, δ) = (22h17m35.0s, +00d17m00.0s) with Nyquist sampling was adopted, which covers 17 of the 18 SMGs in ADF 22 in the area in which the primary ALMA beam response is ≳ 50%. Observations for the 3-mm line scan were performed using between 36 and 44 of the 12-m antennas, under good weather conditions. The observations were divided into 19 individual execution blocks (EBs), and the total on-source time was 13.9 hours. For the phase and bandpass calibrators, PKS 2224+006 and PKS J2148+0657 were observed, respectively. To set the absolute flux scale, Pallas, J PKS J2148+0657, PKS 2230+11, and Titan were observed.

The ALMA data were reduced using the Common Astronomy Software Application (CASA) versions 4.7.0, 4.7.2, and 5.1.0 (*46*). The calibration and flagging were carried out with a standard CASA pipeline. Combining the 19 EBs produced a cube with a velocity resolution of 100 km $s^{-1}$ for each spectral window, using the CASA task tclean, adopting natural weighting and mosaic gridder mode. Continuum components were subtracted using IMCONTSUB. The effective beam size is $0.''93\times 0.''80$ with position angle of $-54$ deg at 84.5 GHz. The data cube has an r.m.s. level of ~89 $\mu$Jy $beam^{-1}$ ($1\sigma$) at the phase-tracking center at 84-86 GHz. We detected CO emission from 14 of the 17 SMGs observed. The one-dimensional spectra extracted within a $2''$ diameter aperture centered on each peak of CO line emission are shown in Fig. S5. A single Gaussian model was fitted to each line to determine the redshifts, listed in Table S2.

# S6 MOSFIRE Data and Analysis

In addition to the ALMA survey described above, we can assign redshifts to the SMGs via rest-frame optical emission lines, which are less sensitive to dust attenuation than rest-frame UV lines. For our targets, which have $z$~3.09, we can observe the H β, [O III] 4959, 5008 Å emission lines in the *K*-band atmospheric window, while the Hα line is not accessible from the ground. We use the Multi-Object Spectrometer for Infra-Red Exploration (MOSFIRE) (*47*) on the Keck-I telescope for *K*-band spectroscopy of the SMGs. The observations were performed during two half nights, 2017 September 29 and 30 as a part of a Subaru-Keck time exchange program (S17B-136 or 2017B_S412, PI. H. Umehata). We used two slit masks with slit widths of 0.7". The integration time was set to 180 sec for each exposure, using a 2-position nod sequence with ±1.5" dithers along slits. The total on-source exposure time was 4.0 hours and 3.2 hours for the two masks, respectively. The data were reduced using the MOSFIRE data reduction pipeline (DRP) (*48)* in a standard manner. One-dimensional spectra were extracted and measured using methods described in detail by previous works (*48, 49*). We observed 11 SMGs, detecting line emission from 10 SMGs at $z$~3.09, of which six had been previously identified (Table. S2 and Fig. S6).

# S7 Cluster Membership and Redshift Distributions of Galaxies

In Fig. S7, we show the spatial distribution of proto-cluster member galaxies in the region of ADF 22, including line-of-sight velocity information. The positions of the proto-cluster galaxies are superposed on the Ly α filaments. Only galaxies with spectroscopic redshifts are considered in this analysis. The galaxies listed include the 15 ALMA-detected SMGs at z=3.09 from the ALMA band 3 and MOSFIRE spectroscopy. We include in the discussion one additional SMG, ADF 22. A10, which is located slightly outside of the MUSE coverage (*32*), because it may contribute to the illumination of the Ly α filament. Fig. S10 shows a histogram of Ly α velocity offsets compared to systemic velocities for SMGs. These are correlated in velocity space, within the typical shifts of ~200-500 km $s^{-1}$ expected for resonant transitions and in presence of inflows and/or outflows (*50*). We have also included the rest-frame UV-to-optical continuum-selected populations, which contain LBGs (*19*, *51-53*) and *K*-band selected galaxies (*54-56*). Together with the archival data and our MOSFIRE observation, seven such galaxies have measured redshifts, in addition to the SMGs. Lastly, previously observed LAEs are also included (*23, 52, 57*), three of which have systemic redshifts measured from nebular emission lines in the rest-frame optical.

# S8 Over-densities of SMGs and X-ray AGNs

The over-densities of SMGs and X-ray AGNs are estimated using published methods (*21*). We consider the projected area of $2' \times 3'$, and a redshift range of $z$=3.08-3.10. There are 5 SMGs with total infrared luminosity $L_{IR}$ >~$10^{12.5}$ $L_{\odot}$ and 8 X-ray AGNs with $L_X$>~$10^{44}$ erg $s^{-1}$ (e.g., *28*, *32*), resulting in volume densities of $\sim 2 \times 10^{-3}$ $Mpc^{-3}$ and $\sim 4 \times 10^{-3}$ $Mpc^{-3}$, respectively. The typical volume densities of the SMGs and X-ray AGNs at $z$~3 are instead $\sim 4 \times 10^{-6}$ $Mpc^{-3}$ (*58*) and $\sim 10^{-5}$ $Mpc^{-3}$ (*59*), implying that SMGs and X-ray AGNs that are three orders of magnitude over dense in the SSA 22 region compared to the general field.

# S9 Radiative Transfer Calculation

The detected surface brightness in the extended filaments exceeds the level expected for fluorescent emission induced by the $z$~3 ultraviolet extragalactic background (UVB) (e.g., *11*). While the majority of the compact emission and the signal associated with the LABs arise from other processes (including scattering and cooling radiation; (*28, 60-62*)), parts of the extended filaments do not appear to be associated with sources above the detection limits of our observations. We use radiative transfer calculations in a simple geometry, to determine whether such isolated emission can arise from fluorescent emission originating from the enhanced radiation field within the SSA 22 region.

We run radiative transfer calculations with the CLOUDY software (*63*), modelling filaments as slabs of constant density. The gas is assumed to have a solar abundance pattern, with an absolute metallicity scaled to 1% of the solar value. This is in line with the abundances of strong absorption line systems *(64*), which we expect is suitable for an over-dense region like SSA 22 . However, we have verified that our calculation is insensitive to the precise value assumed for the metallicity. The slabs are illuminated with an ionizing radiation field with shape similar to the UVB (*9*), which is expected to capture the average radiation field of the SSA 22 region given by a combination of AGNs and star-forming galaxies. On the basis of these assumptions, we run a grid of models, varying the column density of the slab and the intensity of the radiation field, to mimic the local enhancement induced by an over-density of SMGs and AGNs in SSA 22 . Results are shown in Fig. S8. We recover a surface brightness which increases progressively as a function of column density up to the point at which the slab turns optically thick, where the emission plateaus to a value that is solely proportional to the intensity of the ionizing radiation field, in agreement with analytic predictions (*65*).

From the surface brightness variation as a function of column density, it is evident that an enhanced radiation field (> $12 \times$ the UVB) in optically thick gas would match the observed surface brightness, and that an even higher intensity would be required if gas is instead fully ionised. As described in the main text, such an enhancement can be provided by the over-dense SMG/AGN population. In this calculation, we assumed a constant density slab, which is clearly an oversimplifying assumption, as a clumpy medium is indeed often required to model the observed emission, particularly in circumgalactic regions (*12, 66*). Clumpiness does not alter our result in the optically thick regions, where the peak emission is only a function of the radiation field intensity, but it affects the predicted surface brightness in the optically-thin limit. However this does not affect the main conclusion of our modelling.

# S10 Ionizing Luminosity Estimate

The ionizing photon fluxes provided by SMGs and X-ray luminous AGNs are estimated on the basis of the SFR for SMGs and the expected intrinsic spectral energy distribution (SED) in the rest-frame wavelengths covering the extreme UV of the X-ray luminous AGNs. In the case of SMGs, the SFR was calculated from SED template fitting as in previous work (*21*). We fitted the 1.1 mm flux densities measured using ALMA with an SED library of SMGs (*67*), adopting the systemic redshifts derived above. We then calculated a median value of $L_{IR}$, integrating each template SED in the rest-frame 8-1000 micrometres. This $L_{IR}$ was converted to SFR following the Kennicutt relation ( (*68*), their equation 4). We adopted a Salpeter initial mass function (IMF, *69*), although the shape of the IMF of dusty starburst galaxies is still a matter of debate (e.g., *70*). This yielded SFRs in the range of 160-1700 $M_{\odot}$yr$^{-1}$ for the 16 SMGs at $z$=3.09, or a total SFR ~6900 $M_{\odot}$yr$^{-1}$ in the field. The intrinsic ionizing luminosity ($Q_{ion}$) generated from the SMGs are thus $Q_{ion}$ ~$10^{56.9}$ s$^{-1}$ in total, assuming $Q_{ion}$ = $1.08 \times 10^{-53}$ photon s$^{-1}$ per unit $M_{\odot}$ yr$^{-1}$ of SFR ( (*68*), their equation 2). For X-ray AGNs, we estimated the flux at wavelengths shorter than the Lyman limit, following published methods (*28,71*). The X-ray luminosity of the detected X-ray AGNs are on average $L_X$ ~ $10^{44}$ erg s$^{-1}$ (*28*). Scaling the X-ray measurement to match a radio quiet quasar template in the rest-frame (*72*), we derived a monochromatic luminosity at Lyman limit, $\nu_{LL} L_{\nu_{LL}} = 10^{45}$ erg s$^{-1}$, or $L_{\nu_{LL}} = 10^{31.1}$ erg s$^{-1}$ Hz$^{-1}$. We then assumed that the AGN SED obeys a power-law with a form of $L_\nu = L_{\nu_{LL}} (\nu / \nu_{LL})^{-\alpha_Q}$ for $\nu > \nu_{LL}$, where $L_\nu$ is monotonic luminosity at frequency $\nu$, following previous work (*71*). Then, the total ionizing luminosity from the 8 X-ray luminous AGNs was estimated as $Q_{ion}$ ~$10^{57.0}$ s$^{-1}$ via $Q_{ion} = 8 \times \int_{\nu_{LL}}^{\infty} L_\nu / h\nu \, d\nu$ photon s$^{-1}$, adopting a slope of $\alpha_Q$=1.57 (*73*). Therefore the SMGs/AGNs can account for a required photon number rate of $Q_{ion}$ ~$10^{55}$ s$^{-1}$ if the overall escape fraction is order of 1%, which is lower than estimated for typical star-forming galaxies at $z$~3 (*74*).

# S11 Hydrogen Mass Estimate

The mass of hydrogen in the two main filaments are estimated assuming a simple geometry: a 3D cylinder 1 Mpc long and ~100 kpc in diameter. On the basis of radiative transfer calculations (*12*), a typical gas density $6 \times 10^{-3}$ cm$^{-3}$ is adopted if gas is fully ionised. Then, the mass of hydrogen is calculated to be of the order of $\sim 10^{12}$ $M_{\odot}$, an estimate that bears substantial uncertainty due to the unknown thickness of the filaments. This value is directly proportional to the value of an unknown volume filling factor, and thus should be considered only an order of magnitude estimate.

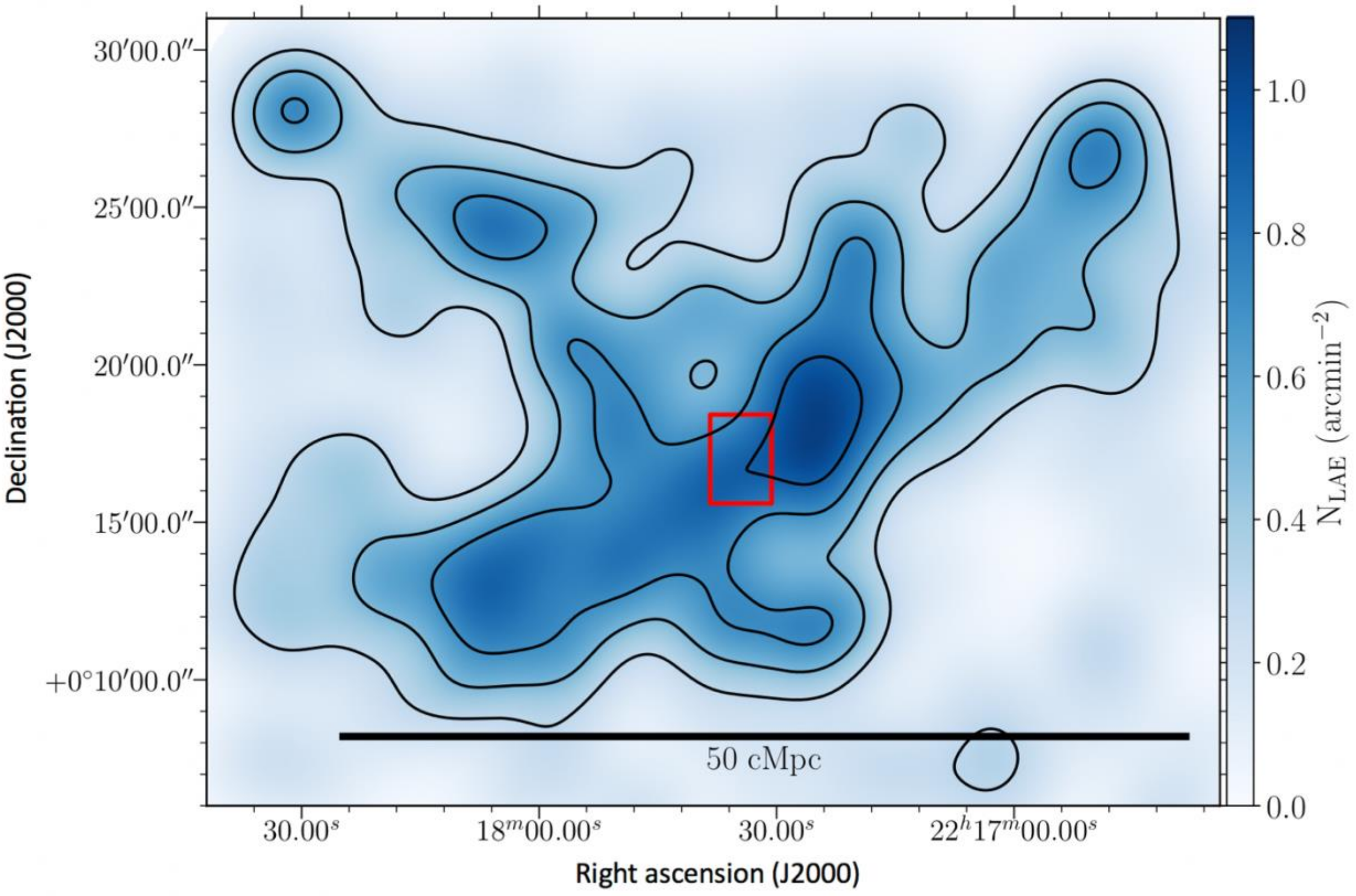

**Fig. S1. The location of ADF 22 compared to projected number density of LAEs at z~3.09**. Here $N_{LAE}$ is the projected number density of LAEs in units of arcmin$^{-2}$ (*31*). ADF 22 is located not near the projected density peak, at a junction of a three-dimensional filamentary structure on a 50-comoving megaparsec scale, traced by the LAEs (a red rectangle, *20*). This indicates that the filaments identified using MUSE reside at the center of the galaxy distribution in the proto-cluster, although how the low surface brightness filaments and large scale structures are associated with each other cannot be investigated with the current data.

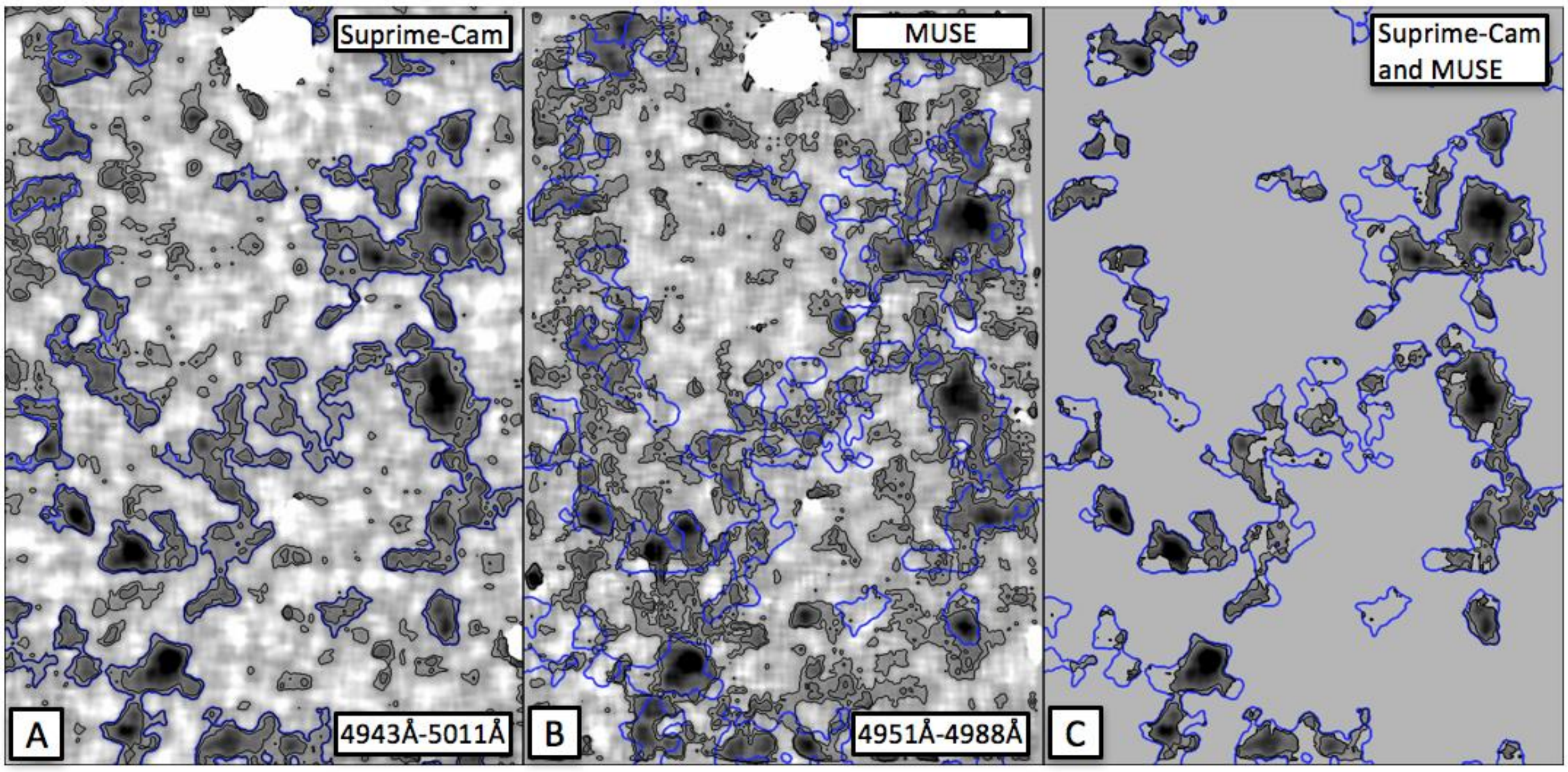

**Fig. S2. Ly α Maps of the target field (116” x 169” in size)**. (A) the median-smoothed narrow-band image taken with Suprime-Cam. Thin contours show the Ly α surface brightness at $(0.3, 0.6) \times 10^{-18}$ erg s$^{-1}$ cm$^{-2}$ arcsec$^{-2}$. Extended Ly α emission extracted with SEXTRACTOR is highlighted with blue thick contours. (B) median-smoothed narrow-band image produced from the MUSE data cube. Contour levels are equivalent to those of the Suprime-Cam image. (C) A subset of the Ly α emission extracted from the Suprime-Cam image in which Ly α emission is also detected in the MUSE narrow-band maps. The filamentary structure is consistent between the two independent observations.

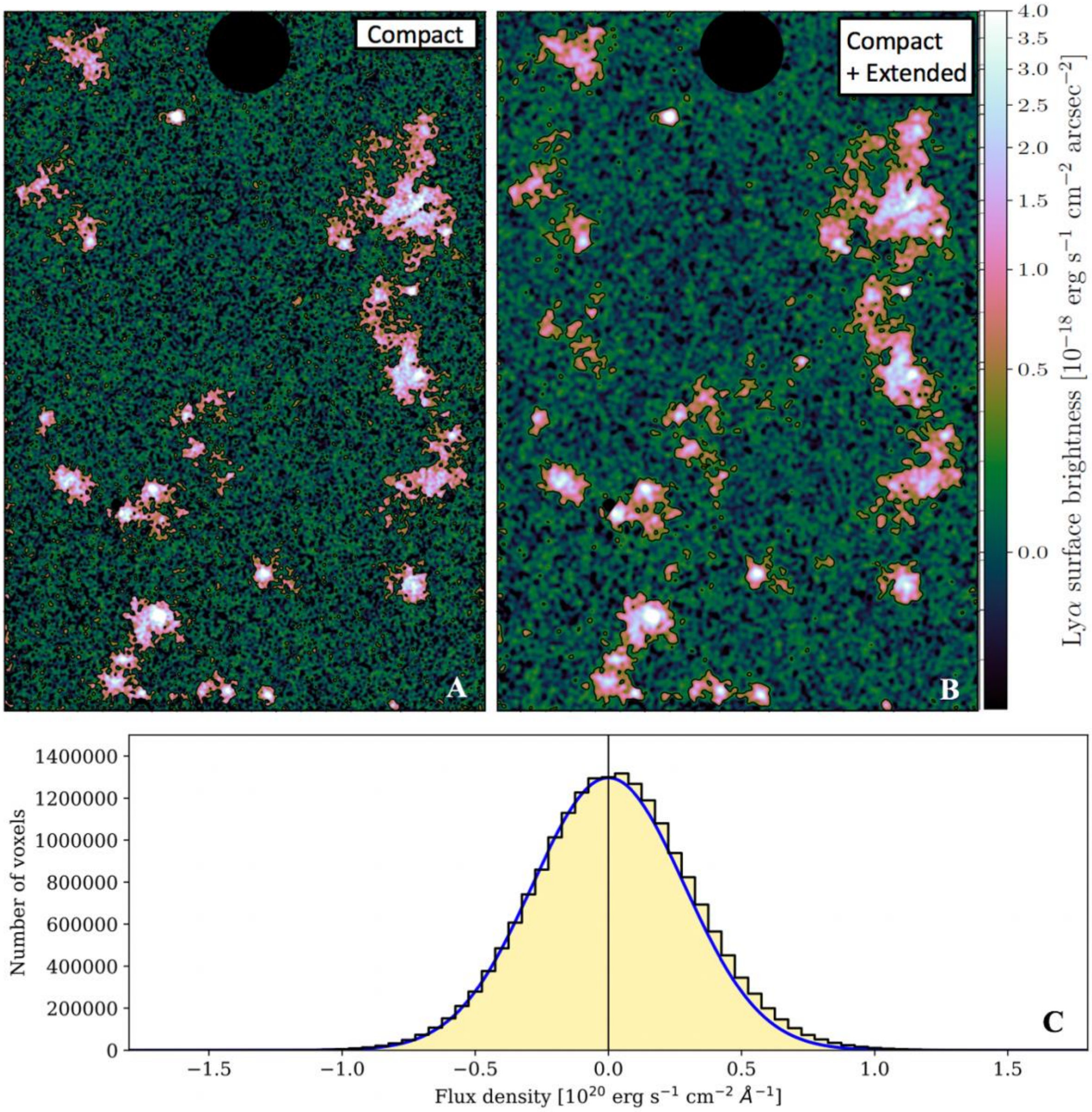

**Fig. S3. MUSE Ly α maps with two layers and flux distributions**. (**A**) A MUSE Ly α map reconstructed with a layer: "compact" emission identified within the MUSE data cube. The image is smoothed with a Gaussian kernel with radius of 3 pixels. Contours represent the Ly α surface brightness at $0.5 \times 10^{-18}$ erg s$^{-1}$ cm$^{-2}$ arcsec$^{-2}$ (corresponding to a $2\sigma$ limit). (**B**) A MUSE Ly α map reconstructed with a "compact + extended" layer which is identified within the MUSE cube at lower surface brightness and also by the Suprime-Cam narrow-band image (see text for details). A Gaussian kernel with radius of 5 pixels is utilized for smoothing, and contours to show a $2\sigma$ limit represent the Ly α surface brightness at $0.3 \times 10^{-18}$ erg s$^{-1}$ cm$^{-2}$ arcsec$^{-2}$. While the majority of the filaments are identified even in the compact layer, some fainter and extended parts appear only in the extended layer. (**C**) The flux distributions of the smoothed MUSE

cube for the “compact+extended” layer (yellow histogram) together with the best fitting Gaussian function for negative fluxes only (blue curve).

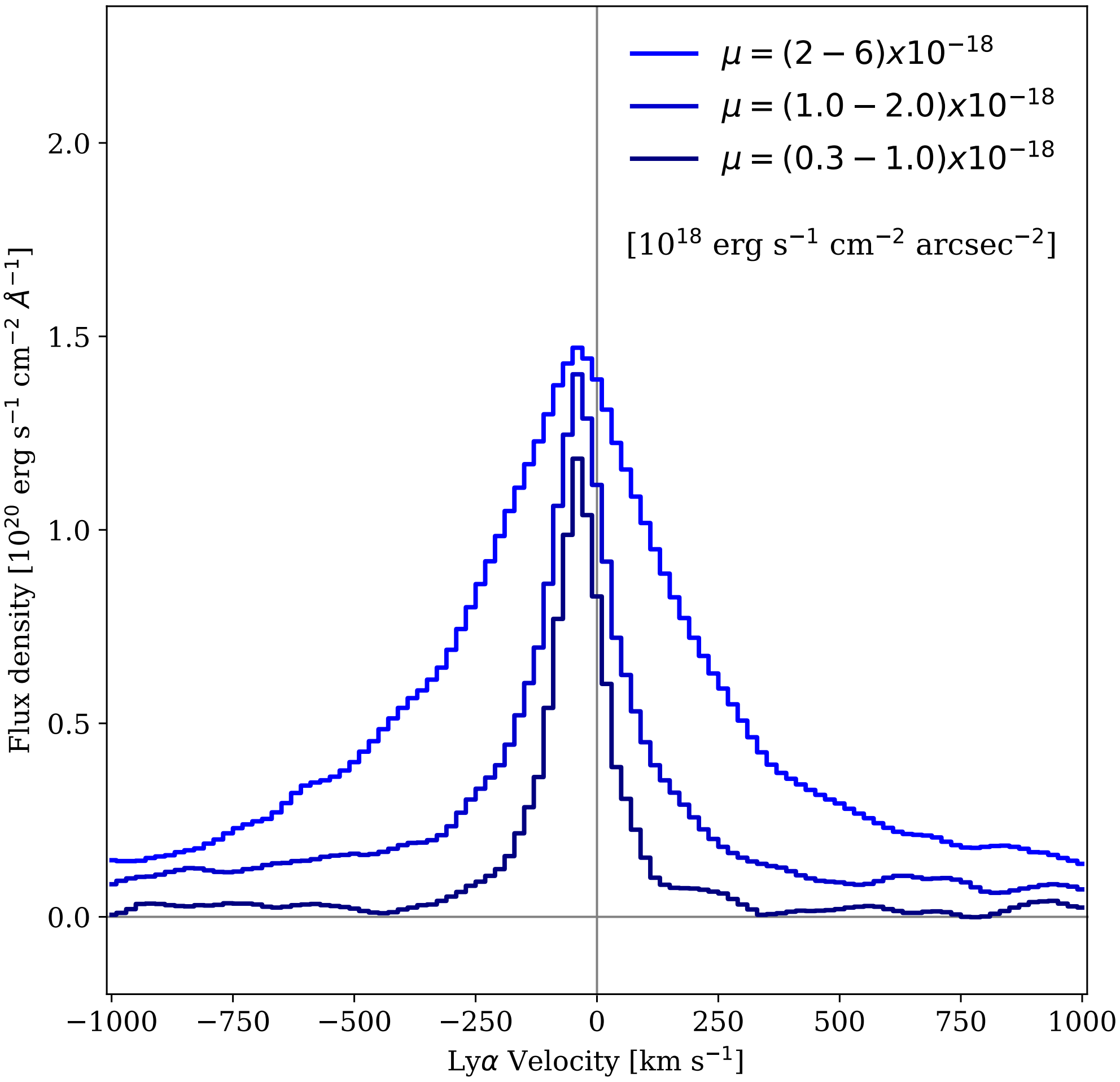

**Fig. S4. Stacked Ly α spectra from the MUSE cube for regions of the filaments with different ranges of Ly α surface brightness.** The fainter parts have on average narrow velocity width than the brighter parts.

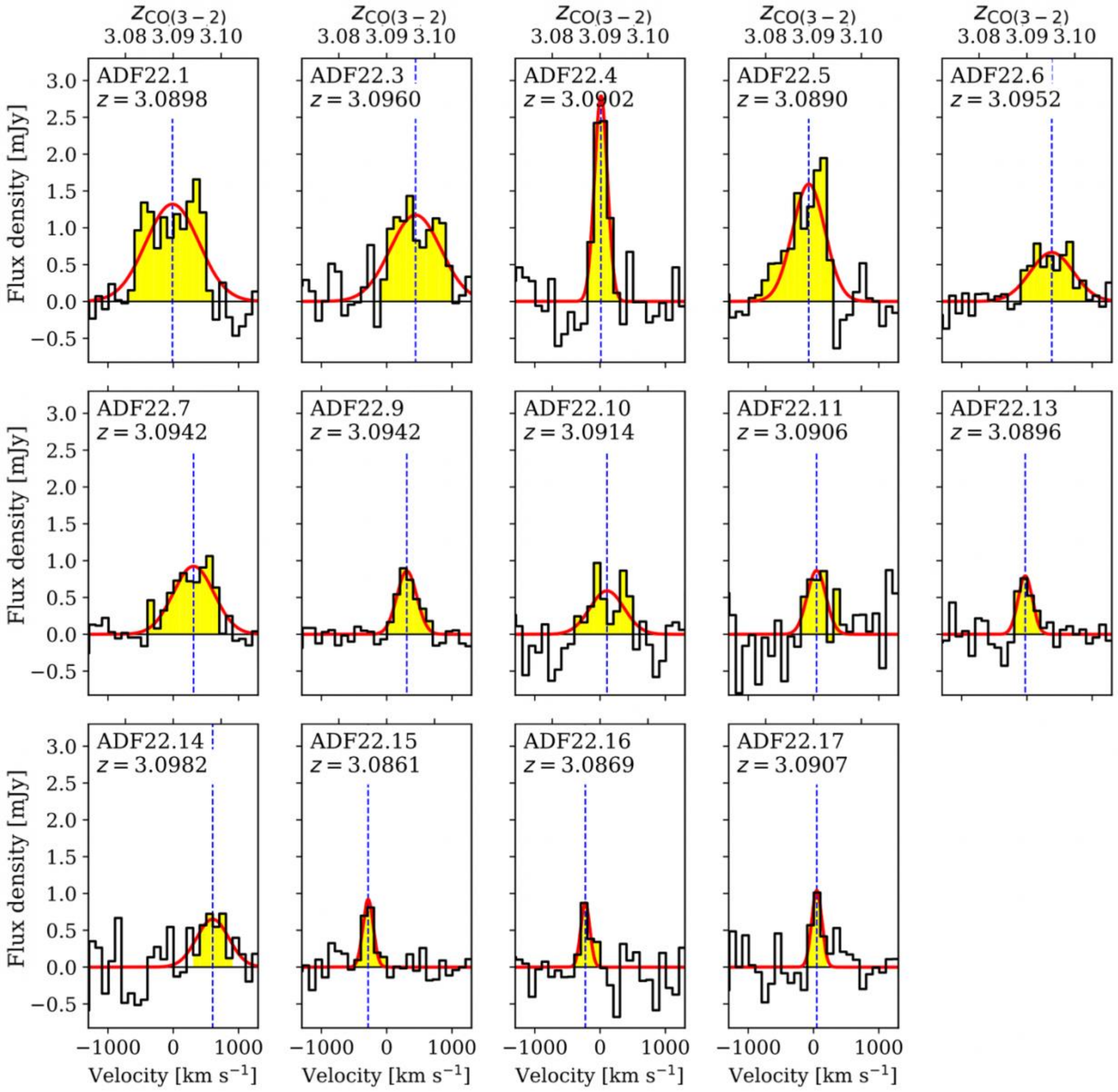

**Fig. S5. The CO ($J$=3→2) line spectra of the SMGs in ADF 22 taken with ALMA band 3**. Red curves and blue vertical lines show the best-fitting Gaussian profiles and the systemic redshifts, respectively.

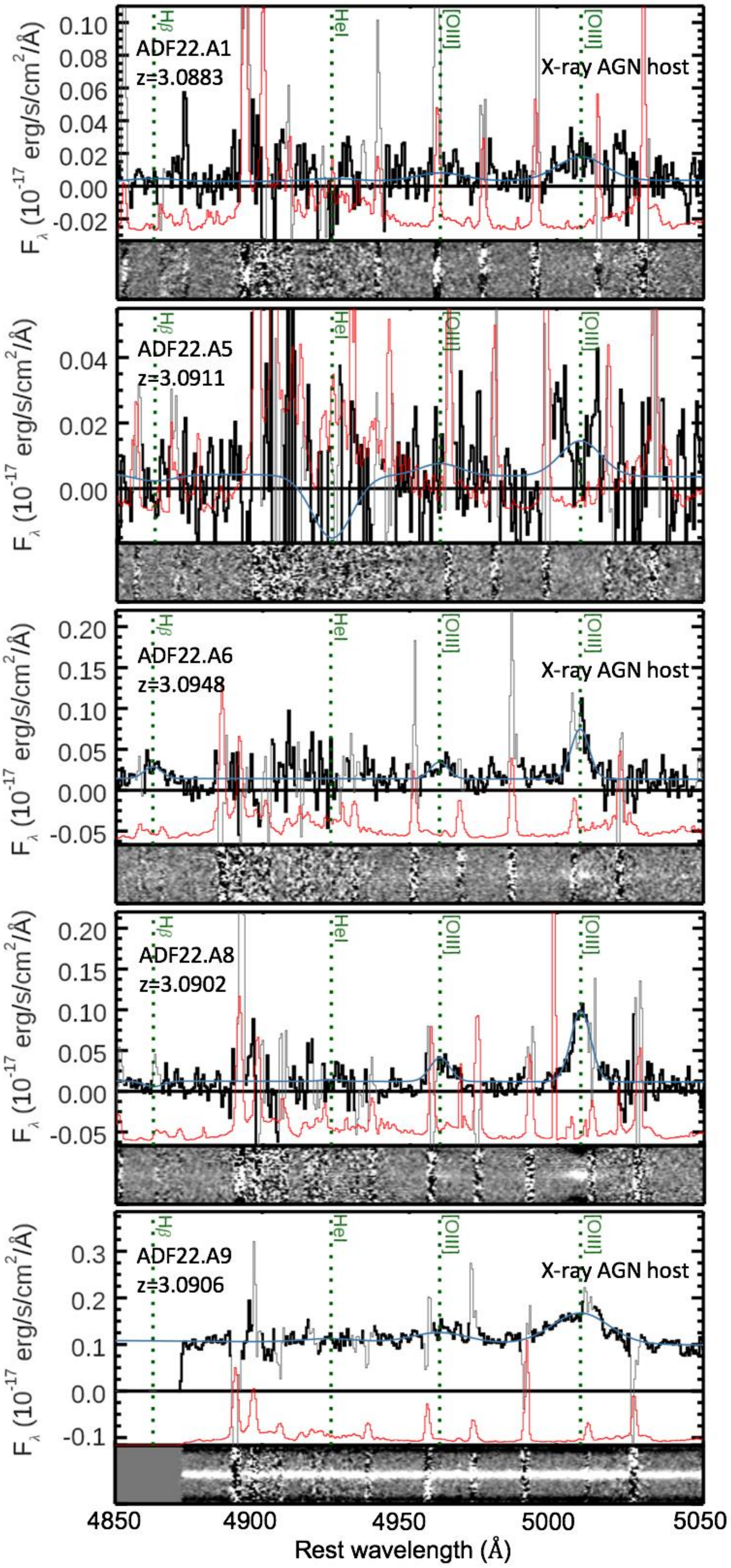

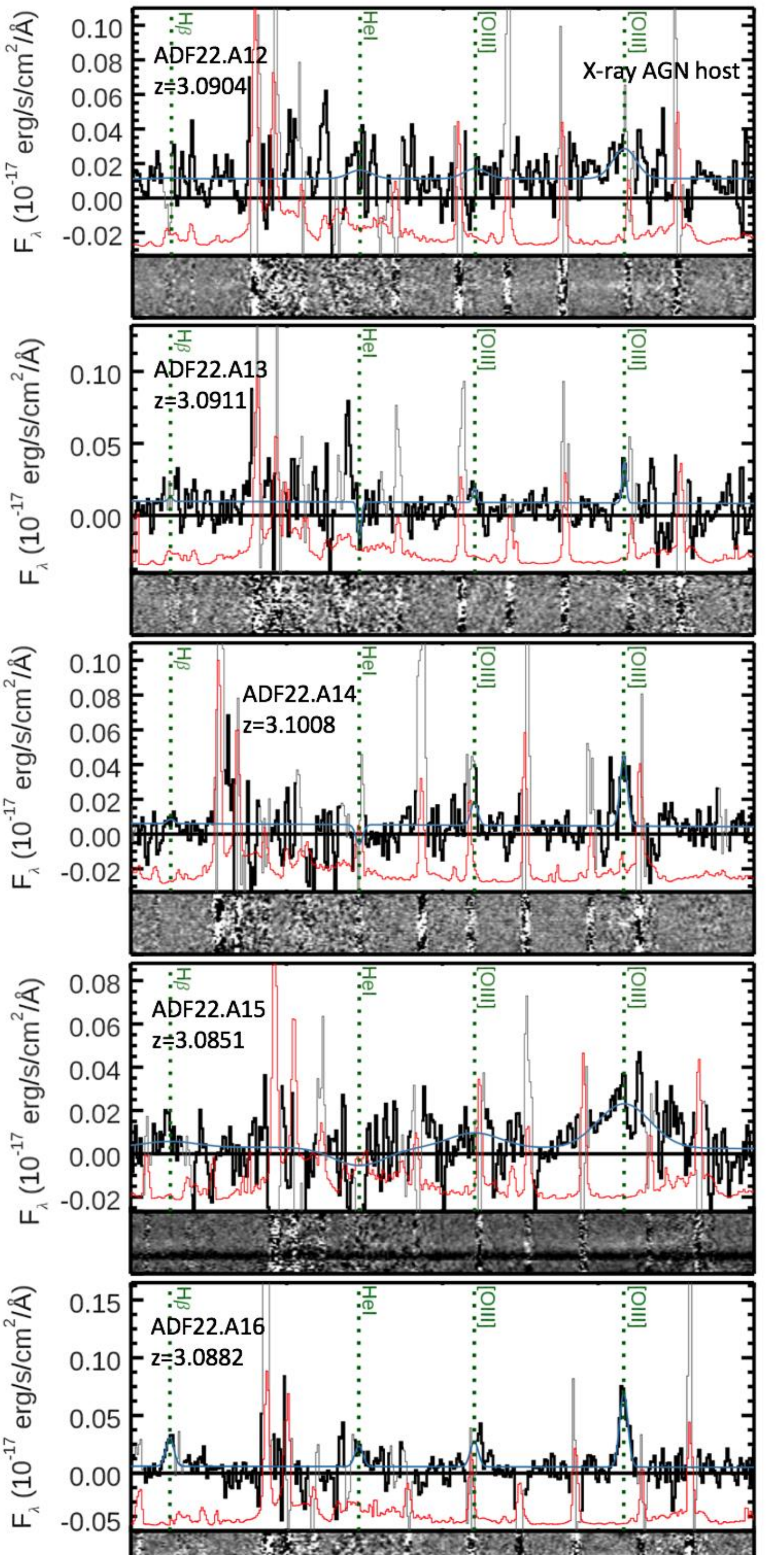

**Fig. S6. *K*-band spectra obtained with Keck/MOSFIRE of the SMGs in ADF 22.** Here we show 10 SMGs with at least one detected emission line, including 4 SMGs host a X-ray AGN. The observed 1-d spectra are shown with solid black lines, while the noisy parts are marked as gray lines. The best-fit line profiles are superposed using a blue line, while the 1 $\sigma$ error spectrum is shown using a red line. The bottom panel shows the stacked two-dimensional spectra from which the 1-d spectra were extracted. Green dotted lines show the rest-frame wavelengths of lines covered in the wavelength range (H β, He I, [O III] 4959, 5008).

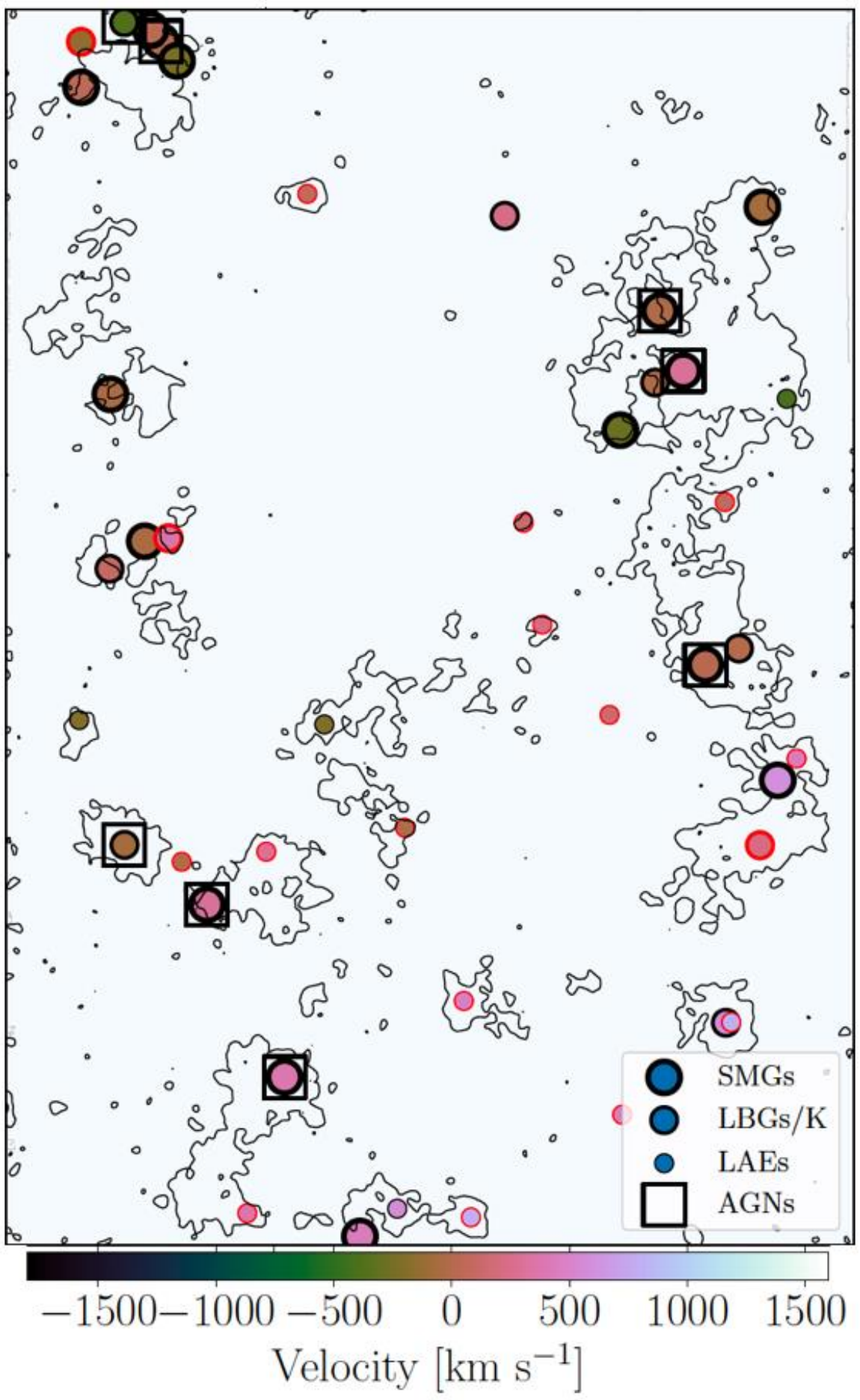

**Fig. S7. Line-of-sight velocity distribution, centered at *z*=3.090, of proto-cluster galaxies.** The SMGs, LBGs, *K*-band selected galaxies, and LAEs, are shown, compared to the Ly α filaments. Symbols, color-coded as in Fig. 3A, are defined in the legend. Symbol edges indicate whether the velocity is determined with CO and/or nebulae lines (black) or only with Ly α emission (red). In addition to SMGs and X-ray AGNs, also galaxies selected in the rest-frame UV and optical appear closely associated with the Ly α filaments.

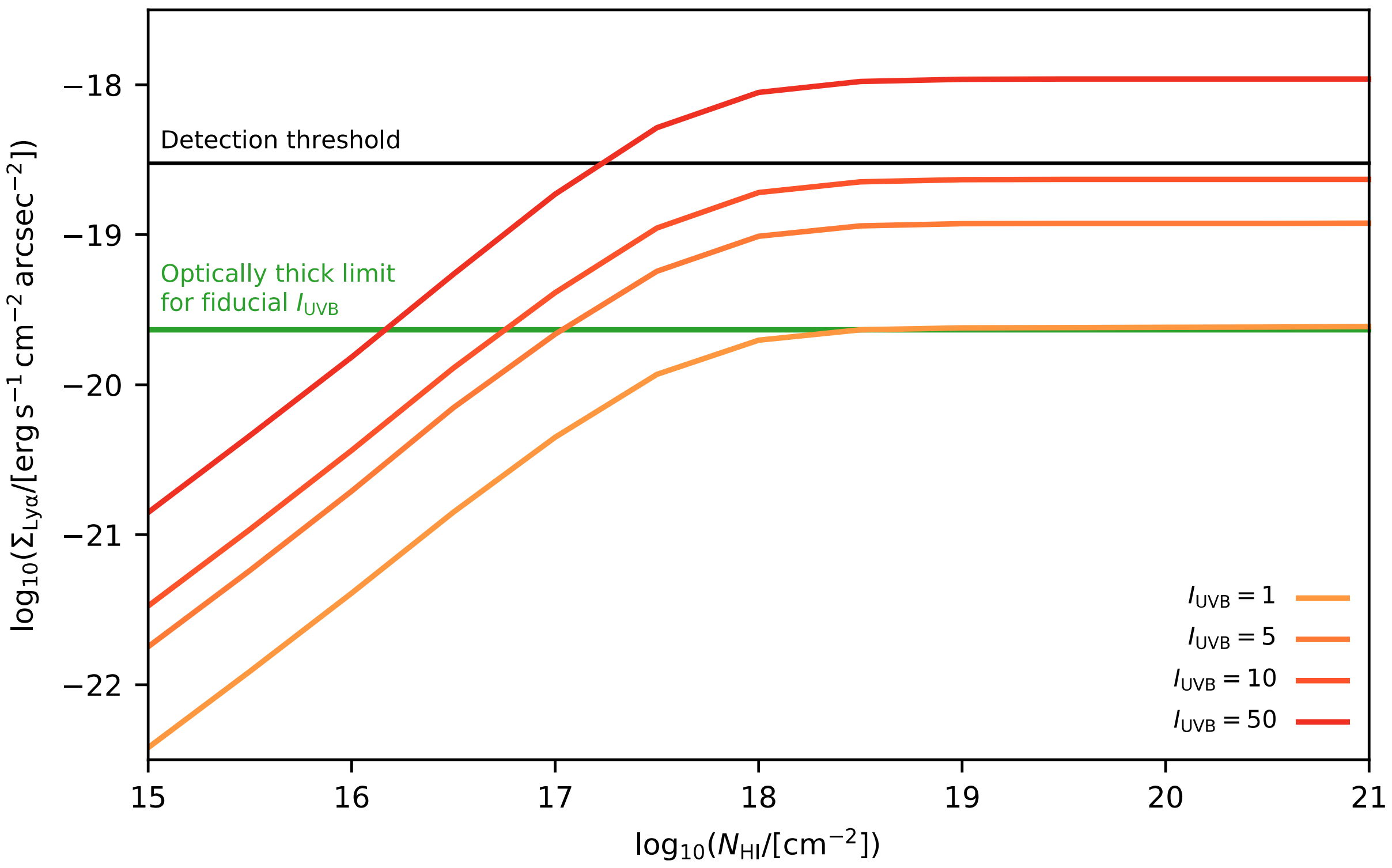

**Fig. S8. Predicted Ly α surface brightness for a slab of varying column density that is illuminated by the ionizing portion of the UVB.** Solid red and orange lines stand for predicted Ly α surface brightness, corresponding to a range of the UVB intensity ($I_{\rm UVB}$) *(9)*, renormalized by a multiplicative scaling factor as in the legend. The green horizontal line shows the analytic solution for emission in photoionized optically-thick gas, while the black line is the detection threshold of the MUSE observations. Elevated UV radiation fields ($> 12\times$ the fiducial UVB) are required to reproduce the Ly α surface brightness in the extended structures observed using MUSE.

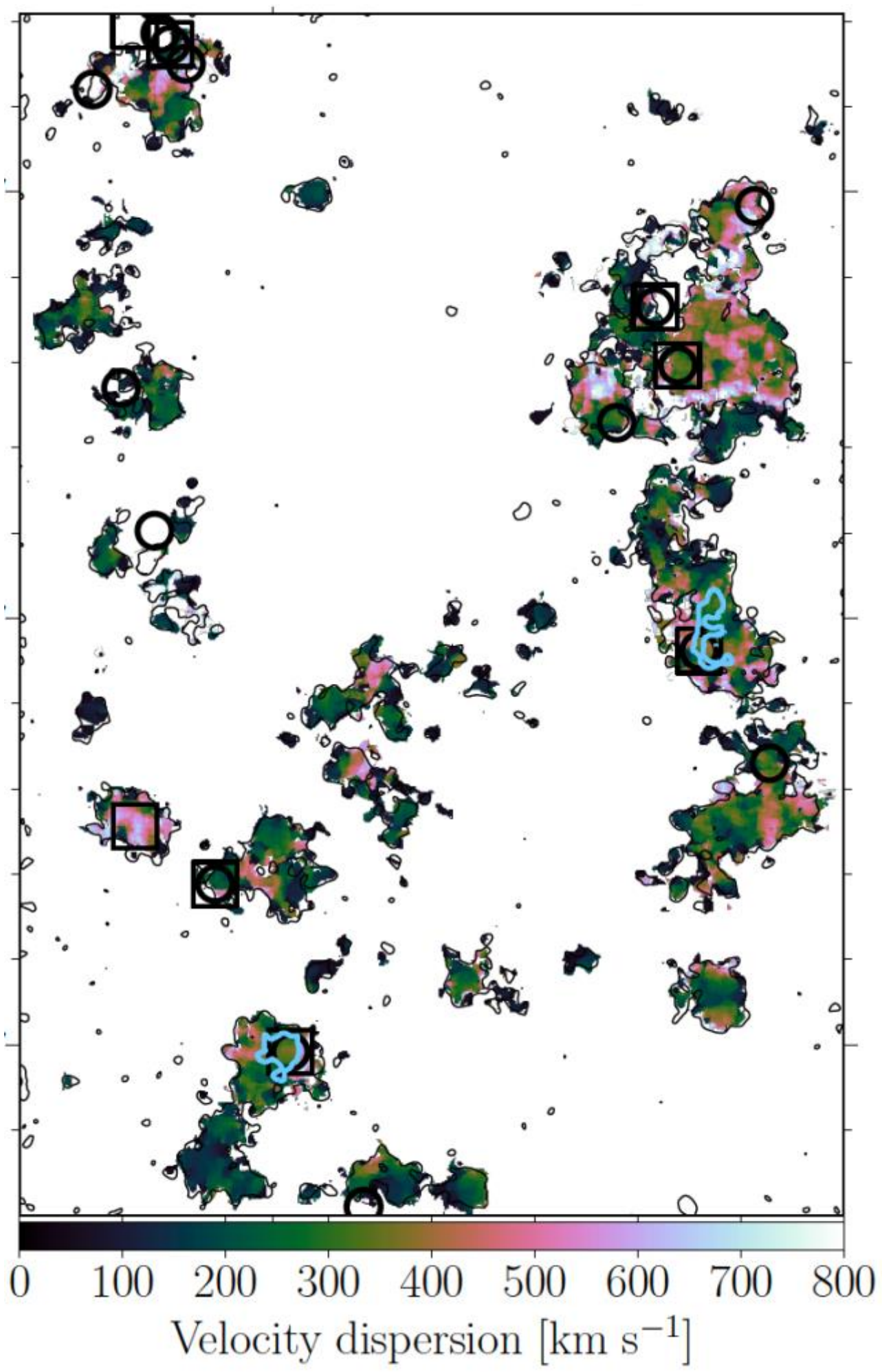

Fig. S9. **Velocity dispersion map created in a same manner as the velocity map of Fig. 3 using CubEx**. Filaments show lower velocity dispersions on average than the regions overlapping with the LABs (cyan contours), which is consistent with what expected for cold gas filaments. Some of patches at large velocity dispersion (~800 km/s) bear a large uncertainty and may be a spurious result from multiple components along the line of sight and low signal-to-noise.

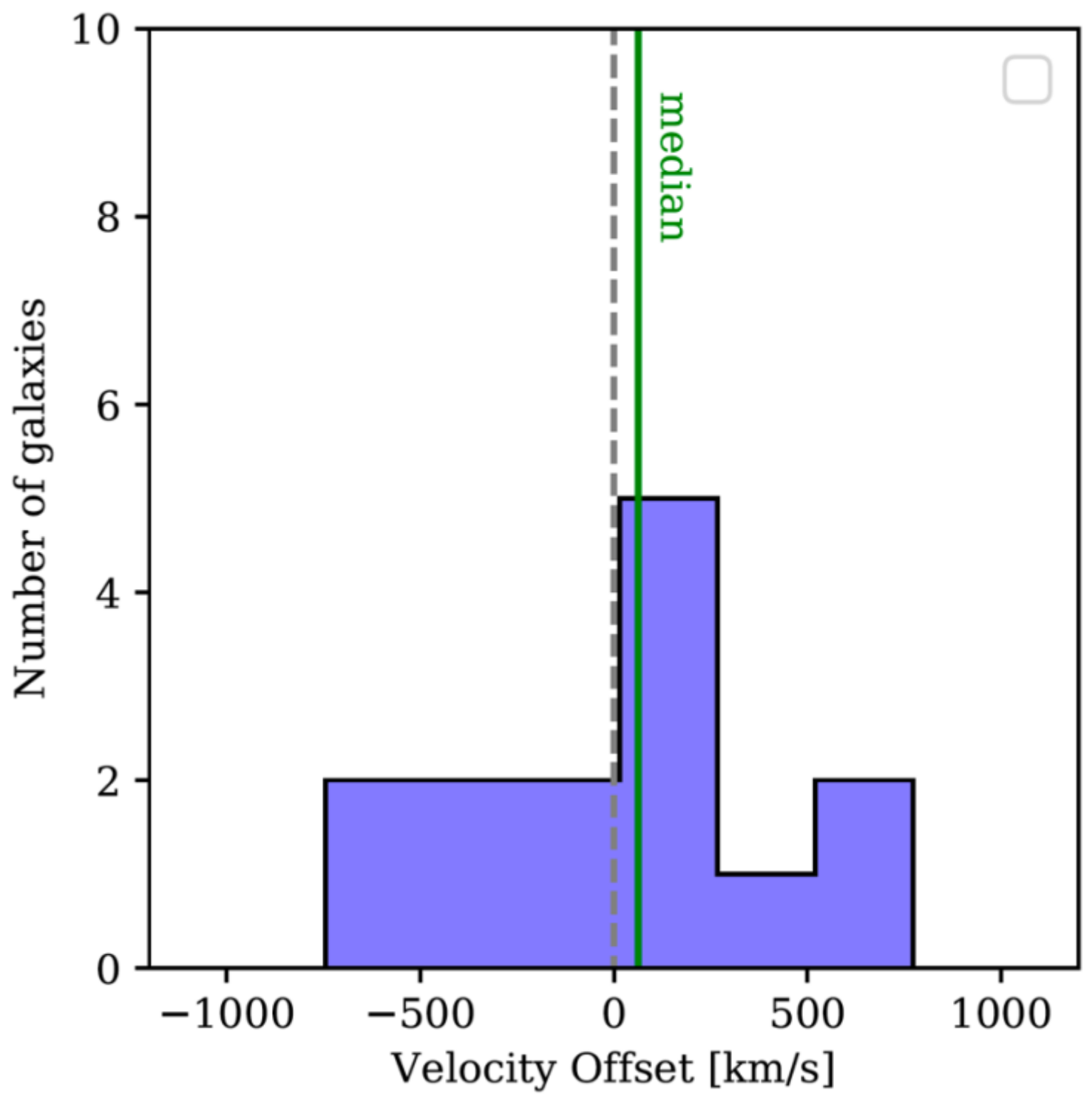

Fig. S10. **Velocity offset between the SMGs/AGNs systemic redshift and the Ly α emission (Ly α velocity – systemic velocity)**. We use systemic velocity (redshifts) derived from CO(*J*=3→2) or [OIII] 5008 Å emissions and flux-weighted Ly α velocity at the position of the galaxies, respectively.

**Table S1. Ly α properties of the *z*=3.09 filaments detected in this work.** We also list the properties of two *z*=3.09 LABs embedded in the filaments (*23*).

| Parameters | Filaments (Compact) | Filaments (Extended) | LAB 12 | LAB 14 |
|---|---|---|---|---|
| Ly α flux [ergs $s^{-1}$ $cm^{-2}$] | $(1.7 \pm 0.4) \times 10^{-15}$ | $(2.3 \pm 0.4) \times 10^{-15}$ | $(1.0 \pm 0.3) \times 10^{-16}$ | $(1.4 \pm 0.3) \times 10^{-16}$ |
| Projected area [$arcsec^2$] | 1441 | 2542 | 29 | 27 |
| Projected area [$kpc^2$] | $8.3 \times 10^4$ | $1.5 \times 10^5$ | $1.7 \times 10^3$ | $1.6 \times 10^3$ |
| Detection Thresholds [ergs $s^{-1}$ $cm^{-2}$ $arcsec^{-2}$] | $5.0 \times 10^{-19}$ | $3.0 \times 10^{-19}$ | $2.2 \times 10^{-18}$ | $2.2 \times 10^{-18}$ |
| Averaged surface brightness [ergs $s^{-1}$ $cm^{-2}$ $arcsec^{-2}$] | $(1.2 \pm 0.3) \times 10^{-18}$ | $(0.9 \pm 0.2) \times 10^{-18}$ | $(3.4 \pm 1.1) \times 10^{-18}$ | $(5.2 \pm 1.1) \times 10^{-18}$ |
| Luminosity [erg $s^{-1}$] | $(1.5 \pm 0.3) \times 10^{44}$ | $(2.0 \pm 0.3) \times 10^{44}$ | $(8.6 \pm 0.3) \times 10^{42}$ | $(1.2 \pm 0.3) \times 10^{43}$ |

**Table S2. Systemic redshifts of the SMGs (from ADF 22. A1 to ADF 22. A17) and AGNs (AGN1, AGN2) at z=3.09 in ADF 22**. Among SMGs, ADF 22. A1, ADF 22. A4, ADF 22. A6, ADF 22. A7, ADF 22.A9, and ADF 22. A12 are a host of a X-ray luminous AGN. The colmuns show ID, Right Ascension, Declination, systemic redshift ($z_{sys}$), Detected lines (we use the first one to determine $z_{sys}$), and References in the literatures.

| ID | Right Ascension (J2000) | Declination (J2000) | $z_{sys}$ | Line(s) | Ref |
|---|---|---|---|---|---|
| ADF 22. A1 | 334.385056 | 0.295500 | 3.0898 ±0.0006 | CO (J=3→2), [O III] | |
| ADF 22. A3 | 334.396445 | 0.260334 | 3.0960 ±0.0007 | CO (J=3→2), [C I] ($^3P_1 \rightarrow {}^3P_0$) | *(40,75)* |
| ADF 22. A4 | 334.404000 | 0.305750 | 3.0902 ±0.0001 | CO (J=3→2), CO (J=9→8) | *(76)* |
| ADF 22. A5 | 334.381167 | 0.299445 | 3.0890 ±0.0005 | CO (J=3→2), [O III] | |
| ADF 22. A6 | 334.399306 | 0.266389 | 3.0952 ±0.0008 | CO (J=3→2), [O III] | |
| ADF 22. A7 | 334.384167 | 0.293223 | 3.0942 ±0.0006 | CO (J=3→2) | |
| ADF 22. A8 | 334.404611 | 0.286750 | 3.0902 ±0.0004 | [O III] | *(48)* |
| ADF 22. A9 | 334.402250 | 0.272945 | 3.0942 ±0.0003 | CO (J=3→2), [O III] | |
| ADF 22. A10 | 334.404584 | 0.307445 | 3.0914 ±0.0013 | CO (J=3→2) | |
| ADF 22. A11 | 334.404389 | 0.306195 | 3.0906 ±0.0008 | CO (J=3→2), [O III] | *(48)* |
| ADF 22. A12 | 334.383333 | 0.282056 | 3.0904 ±0.0014 | [O III] | *(48)* |
| ADF 22. A13 | 334.405917 | 0.292334 | 3.0896 ±0.0004 | CO (J=3→2), [O III] | |
| ADF 22. A14 | 334.380583 | 0.277667 | 3.0982 ±0.0010 | CO (J=3→2), [O III] | |
| ADF 22. A15 | 334.386556 | 0.290973 | 3.0861 ±0.0002 | CO (J=3→2), [O III] | |
| ADF 22. A16 | 334.403389 | 0.305000 | 3.0869 ±0.0004 | CO (J=3→2), [O III] | *(48)* |
| ADF 22. A17 | 334.407028 | 0.304000 | 3.0907 ±0.0004 | CO (J=3→2) | |
| AGN1 | 334.405380 | 0.275212 | 3.089 | [O III] | *(48)* |
| AGN2 | 334.405390 | 0.306470 | 3.085 | [O III], H $\beta$ | *(48)* |